\documentclass[%
reprint,
nofootinbib,
amsmath,amssymb,
aps,
floatfix,
10pt
]{revtex4-2} 
\usepackage[latin1]{inputenc}
\usepackage{amsmath,amssymb}
\usepackage{mathrsfs} 
\usepackage[capitalise]{cleveref}
\usepackage{siunitx}
\usepackage{braket}
\usepackage{calc}
\usepackage{tabularx}
\usepackage{dsfont}
\usepackage{color}
\usepackage{seqsplit}
\usepackage{ifthen}
\usepackage[normalem]{ulem}
\usepackage{graphicx}
\usepackage{mathtools} 

\allowdisplaybreaks

\newcommand\bignorm[1]{\bigg\lVert#1\bigg\rVert}

\newcommand{\Pe}[0]{{P\'eclet}}

\newcommand{\hparamvalue}[0]{u}

\newcommand{\kB}[0]{{k_\mathrm{B}}}
\newcommand{\tlj}[0]{{\tau_{\mathrm{LJ}}}}
\newcommand{\stddev}[0]{\varsigma}

\newcommand{\myev}[1]{{\widehat{\boldsymbol{#1}}}}
\newcommand{\mymat}[1]{{\underline{#1}}}
\newcommand{\myvec}[1]{{\boldsymbol{#1}}}

\newcommand{\lammps}[0]{LAMMPS}
\newcommand{\dif}{\mathrm{d}}%

\newcommand{\norm}[1]{\lVert#1\rVert}%

\newcommand{\ZT}[1]{\textquotedblleft#1\textquotedblright}%

\newcolumntype{Y}{>{\centering\arraybackslash}X}%
\newcolumntype{Z}{>{\raggedright\arraybackslash}X}%

\newlength{\myl}%
\newcommand{\SUM}[2]{{\setlength{\myl}{\widthof{$\displaystyle\sum_{#1}^{#2}$}*\real{0.5}-\widthof{$\displaystyle\sum$}*\real{0.5}}\sum_{#1}^{#2}\;\hspace{-\the\myl}}}
\newcommand{\INT}[3]{\settowidth{\myl}{$\displaystyle\int_{#1}^{#2}$}{\int_{#1}^{#2}\;\;\;\hspace{-\the\myl}\dif #3}\,}
\newcommand{\TINT}[3]{\settowidth{\myl}{$\int_{#1}^{#2}$}{\int_{#1}^{#2}\!\ifthenelse{\equal{#1#2}{}}{}{\;\;\;\;\hspace{-\the\myl}}\dif #3}\,}%
\newcommand{\EINT}[3]{\settowidth{\myl}{$\int_{#1}^{#2}$}{\int_{#1}^{#2}\;\;\;\,\hspace{-\the\myl}\dif #3}\,}

\begin{document}
\title{Collective dynamics and pair-distribution function of active Brownian ellipsoids}

\author{Stephan Br\"oker}
\affiliation{Institute of Theoretical Physics, Center for Soft Nanoscience, University of M\"unster, 48149 M\"unster, Germany}

\author{Michael te Vrugt}
\affiliation{Institute of Theoretical Physics, Center for Soft Nanoscience, University of M\"unster, 48149 M\"unster, Germany}

\author{Raphael Wittkowski}
\email[Corresponding author: ]{raphael.wittkowski@uni-muenster.de}
\affiliation{Institute of Theoretical Physics, Center for Soft Nanoscience, University of M\"unster, 48149 M\"unster, Germany}

\begin{abstract}
While the collective dynamics of spherical active Brownian particles is relatively well understood by now, the much more complex dynamics of nonspherical active particles still raises interesting open questions. Previous work has shown that the dynamics of rod-like or ellipsoidal active particles can differ significantly from that of spherical ones. Here, we obtain the full state diagram of active Brownian ellipsoids depending on the \Pe{} number and packing density via computer simulations. The system is found to exhibit a rich state behavior that includes cluster formation, local polar order, polar flocks, and disordered states. Moreover, we obtain numerical results and an analytical representation for the pair-distribution function of active ellipsoids. This function provides useful quantitative insights into the collective behavior of active particles with lower symmetry and has potential applications in the development of predictive theoretical models. 
\end{abstract}
\maketitle

\section{Introduction}
The physics of active matter \cite{MarchettiJRLPMS2013hydrodynamics,bechinger2016active}, which consists of self-propelled particles, is one of the central areas of soft matter physics. Of particular interest in this context is the collective dynamics of active particles. Here, a variety of phenomena have been studied, most notably \textit{motility-induced phase separation} (MIPS) \cite{Cates15mips}, which is the spontaneous formation of a high-density and a low-density phase in a system of repulsively interacting particles, and \textit{flocking} \cite{Vicsek1995,TonerTu1995long,TonerTR2005hydrodynamics}, which refers to the coherent collective motion of active particles. Both phenomena continue to be widely investigated today \cite{PaoluzziLP2022,OmarRMB2023,AndersonF2022,KreienkampK2022,YuT2022,CapriniL2023}.

The collective dynamics of active matter becomes considerably more complex if the particles are not -- as assumed in many theoretical studies -- spherical. For instance, studies of active ellipsoids \cite{Gromann2020} have found that ellipsoidal particle shapes suppress MIPS and give rise to a rich state diagram involving polar and nematic phases. Moreover, elongated particle shapes allow to study how phenomena known from classical liquid crystal physics, such as topological defects, are modified in active systems \cite{AroraSG2022}. Consequently, the study of the collective dynamics of active particles with interaction potentials that have no spherical symmetry has been a growing field of research in the past years \cite{RebochoTD2022,Gromann2020,JayaramFS2020,vanderlindenAAC2019interrupted,SumaGMO2014,LiaoHK2020,SeseLLPK2022,AroraSG2022,vanDamme19, TheersWQWG2018clustering, ShiC2018self,Fan2017}. See Refs.\ \cite{BaerGHP2020rodReview,Wensink2013,Chate2020} for reviews.

When obtaining the state diagram of active spheres, one usually considers the dependence on the \Pe{} number $\mathrm{Pe}$ (measuring the activity and temperature) and the packing density $\Phi_0$ \cite{Jeggle2020,TakatoriSB2015}. These parameters are also used in state diagrams for active ellipsoids \cite{AroraSG2022}, although previous work has focused more on studying the effects of the particle shape \cite{PeruaniDB2006nonequilibrium,WensinkL2012Emergent,vanDamme19,TheersWQWG2018clustering, ShiC2018self,Fan2017}. Consequently, to understand the collective dynamics of active ellipsoids and to connect it to what is already known about active spheres, more detailed investigations of the state diagram of active ellipsoids as a function of $\mathrm{Pe}$ and $\Phi_0$ are required.

A particularly useful quantity for understanding the collective dynamics of simple and complex fluids is the pair-distribution function $g$ \cite{HansenMD2009}, which determines how the position (and possibly orientation) of a particle depends on that of a reference particle. This function is important because it allows to calculate thermodynamic properties of a system and because it appears in microscopically derived field theories \cite{teVrugtBW2022,BickmannW2020,BickmannW2020b,BickmannBJW2021chirality,broeker2022orientation,teVrugtFHHTW2022,BickmannBroekertVW2022external}. Field-theoretical models are now also being developed for nonspherical active particles \cite{Gromann2020,JayaramFS2020,Bickmann2022Diss,teVrugtHKWT2022}, making knowledge of the pair-distribution function for such particles desirable. In a passive fluid, the pair-distribution function can be studied using liquid integral theory \cite{HansenMD2009}. Methods of this type are, however, not applicable in active systems which are far from equilibrium. Therefore, the pair-distribution function of active particles is less well understood than that of passive ones. 

Previous work on the pair-distribution function of active systems has focused on the case of spherical particles. This was investigated for both single-component systems \cite{BialkeLS2013} and multi-component systems \cite{WittkowskiSC2017}, in both two \cite{Jeggle2020} and three \cite{Broeker2020} spatial dimensions. \citet{HaertelRS2018} have studied both the two-dimensional pair-distribution function and the three-body distribution, and \citet{SchwarzendahlM2018} considered effects of hydrodynamic interactions. What is still missing, however, is a fully orientation-resolved pair-distribution function -- as obtained for spherical particles in Refs.\ \cite{Jeggle2020,Broeker2020} -- for systems of active ellipsoids.

In this article, we significantly extend the results discussed above by systematically investigating the collective dynamics of active ellipsoids via computer simulations. This allows to obtain the state diagram as a function of $\mathrm{Pe}$ and $\Phi_0$, which complements previous works focusing more on the influence of the particle shape and allows for an easier comparison to state diagrams for active spheres (which usually focus on the influence of $\mathrm{Pe}$ and $\Phi_0$). A quantitative analysis allows to classify the observed states into cluster states, local polar order, polar flocks, and global disorder. Moreover, we numerically investigate the pair-distribution function of the active ellipsoids, discuss in detail its symmetries and dependencies on distance and orientations, and obtain an analytical representation that can be used in further theoretical work.

This article is structured as follows: In Section \ref{section:State_diagram_ellipsoids}, we study the state diagram. The pair-distribution function is investigated in Section \ref{section:3_Pair_correlation_function_ellipsoids}. We conclude in Section \ref{sec:conclusion}. In the Appendix, we provide details on the particle interactions (Appendix \ref{sec:interactionpotential}), the order parameters (Appendix \ref{sec:orderparameters}), and the fit parameters (Appendix \ref{sec:tables}).

\section{\label{section:State_diagram_ellipsoids}State diagram}
We start by numerically investigating the state diagram of active ellipsoids. 

\subsection{State of the art}
The state diagram of active ellipsoids is more complex than that of standard active Brownian particles (ABPs) since the anisotropic particle shapes allow for torques. This can lead to polarization and the emergence of nematic phases~\cite{AdhyapakRT2013live,ShankarSBMV2022topological}.
Typically, MIPS is suppressed in systems of active ellipsoids compared to active spheres due to the torque interactions~\cite{Gromann2020, JayaramFS2020}. When undergoing MIPS, spherical particles collide and hinder each other's movement.
If this happens to multiple particles at once, an aggregate can form that other particles then collide with. Initially, particles on the outer layer swim towards the center of the newly forming cluster and exert an inward pressure onto the surface, thereby stabilizing the cluster~\cite{TailleurCates08, Cates15mips}. After some time, their orientations can change due to rotational diffusion, but other particles have already formed a new outer layer that adds pressure and traps the particles in the cluster.

The torque resulting from particle-particle interaction can suppress aggregation at an existing cluster's surface for nonspherical particles such as rods and ellipsoids moving along their long axes.
An active ellipsoidal particle that is oriented roughly, but not perfectly, towards the center of a cluster will experience a torque that turns the particle's orientation away from the cluster's center and thereby prevents it from swimming towards there.
This effect suppresses the cluster aggregation. Even minor anisotropies, like a length-to-diameter ratio of an elliptic particle such as $\kappa= a_\mathrm{el}/b_\mathrm{el} = 1.0424$ (with the length $a_\mathrm{el}$, the diameter $b_\mathrm{el}$, and the length-to-diameter ratio $\kappa$) can cause MIPS aggregates to dissolve~\cite{Gromann2020}.

However, torque can also enhance cluster formation.
This effect was reported by \citet{ZhangAYWG2021active}, who used spherical particles with an anisotropic interaction where repulsion is stronger at the rear of the particles than at the front. This leads to a torque turning the particles towards high-density areas, which enhances MIPS.
While the torque resulting from the shape interaction inhibits MIPS, it can lead to the formation of other phases. These were studied in recent work by \citet{Gromann2020} and \citet{JayaramFS2020}.
\citet{Gromann2020} started with spherical particles undergoing MIPS and then changed the shape of the particles. It was found that even a cluster of spherical particles that has already formed can melt if the shape becomes slightly anisotropic. Increasing the length-to-diameter ratio further can lead to the emergence of local nematic order while the system is in global disorder. For even higher ratios (such as $\kappa = 1.96$), highly polarized domains with local nematic order emerge.
Finally, extremely high anisotropies allow for the formation of a polar band with local smectic order. Similar findings were reported by \citet{PeruaniDB2006nonequilibrium}, who studied hard rods in an overall low-packing density environment and found highly polarized swarms.
\citet{JayaramFS2020}, who studied self-propelled ellipsoids, found that MIPS clusters become polar domains and polar bands for increasing anisotropies in the case of infinite \Pe{} numbers.
The force imbalance coefficient -- an order parameter that characterizes the asymmetry of the interaction force of a particle, an asymmetry that is the cause of MIPS -- reduces for increasing anisotropies. Due to their movement, active particles collide; therefore, particle interaction hinders their movement.
If the particles are elongated, they can slide past each other more easily and do not hinder each other's movement that much, which reduces the trapping effect that causes phase separation~\cite{vanDamme19}. Rather than a persistent large single cluster as in the case of active spheres, one therefore observes a continuous appearance and disappearance of small clusters. This happens even for small anisotropies such as $\kappa = 1.15$ and in the absence of diffusion (infinite \Pe{} number). Further increasing the anisotropy, another state transition occurs, and a single, highly polarized cluster emerges. 
Also, recent experimental findings support that inter-particle torque and alignment of particles hinder MIPS and break full phase separation~\cite{vanderlindenAAC2019interrupted}. 

Even though we consider a wide range of \Pe{} numbers in our work, the active motion is dominant in all cases here. 
For substantially lower \Pe{} numbers (not considered here), the Brownian motion dominates and the collective dynamics differs. Nematic and smectic phases can also emerge \cite{BottWMSBW2018}. For self-propelled rods at infinite \Pe{} numbers (also not considered here), dynamical states such as swarming,
turbulence, and jamming can be found~\cite{WensinkL2012Emergent}.

\subsection{Equations of motion}
The general equations of motion for uniaxial ABPs in two spatial dimensions that can be influenced by interactions, external fields, and orientation-dependent propulsion read~\cite{tenHagenTL2011}
\begin{align}
    \dot{\myvec{r}}_i &= 
    \frac{\mymat{D}_\mathrm{T}}{k_\mathrm{B} T}
    \Big(F_\mathrm{A}\myev{u}_i 
    + \myvec{F}_{\mathrm{int},i} (\lbrace \myvec{r}_j,  \myev{u}_j \rbrace   ) 
    + \myvec{F}_{\mathrm{ext}}(\myvec{r}_i)
    \Big)
    + \myvec{\xi}_{\mathrm{T}} \, , \label{eq:dotr}\\ 
    \dot{\varphi}_i &= \frac{D_\mathrm{R}}{\kB T} M_{\mathrm{int}, i} (\lbrace \myvec{r}_j,  \myev{u}_j \rbrace   )  + {\xi}_{\mathrm{R}}
    \label{eq:dotphi}
\end{align}
with the center of mass position of the $i$-th particle $\myvec{r}_i$, its orientation $\varphi_i$, the translational diffusion matrix $\mymat{D}_\mathrm{T}$, the Boltzmann constant $k_\mathrm{B}$, the temperature $T$, the magnitude of the active propulsion force $F_\mathrm{A}$, the normalized orientation vector $\myev{u}_i$ corresponding to $\varphi_i$, the interaction force $\myvec{F}_{\mathrm{int},i}$ acting on the $i$-th particle where $\lbrace \myvec{r}_i, \myev{u}_i \rbrace$ denotes the set of all particles' positions $\myvec{r}_j$ and orientations $ \myev{u}_j $,
the external force $ \myvec{F}_{\mathrm{ext}}(\myvec{r}_i)$, the translational Brownian noise $ \myvec{\xi}_{\mathrm{T}}$, the rotational diffusion coefficient $D_\mathrm{R}$, the torque $M_{\mathrm{int}, i}$ acting on the $i$-th particle resulting from interactions with other particles, and the rotational Brownian noise $ {\xi}_{\mathrm{R}}$. We specify the interaction forces and torques in Appendix \ref{sec:interactionpotential}.

As ellipsoids are anisotropic, their translational Brownian motion is also anisotropic~\cite{HanANZLY2006}.
The translational Brownian noise $\myvec{\xi}_{\mathrm{T}}$ is implemented by a random force with zero mean.
The variances of its components $k$ and $l$ are given by $\langle {\xi}_{\mathrm{T}, k}(t), {\xi}_{\mathrm{T}, l}(t^\prime) \rangle = 2 {D}_{\mathrm{T},kl} (\varphi (t)) \delta(t-t^\prime)$ where $k,l\in\lbrace x, y \rbrace$ and $\delta(t-t^\prime)$ is the Dirac delta function.
This approach has already been utilized for passive particles in Ref.~\cite{HanANZLY2006} and active particles in Ref.~\cite{tenHagenTL2011}.
Similarly, the rotational Brownian noise is a random torque with zero mean, which variance is given by $\langle \xi_{\mathrm{R}}(t), \xi_{\mathrm{R}}(t^\prime) \rangle = 2 D_\mathrm{R}  \delta(t-t^\prime) $.

In this work, we focus on ellipsoids of revolution that have the same volume as a sphere with diameter $\sigma_0$. The length of their axis of revolution, which is also referred to as the polar axis, is $a_\mathrm{el} = 4^{1/3}\, \sigma_0 $.
The axes perpendicular to the polar axis, which will be referred to as the equatorial axes, have a length of $b_\mathrm{el} = 2^{-1/3} \, \sigma_0$. This means that the ellipsoids have a length-to-diameter ratio $\kappa=2$ and the same volume as a sphere of the same diameter.
An ellipsoid's orientation is defined along its polar axis, which is also the direction of the propulsion force.
Note that $a_\mathrm{el}$ and $b_\mathrm{el}$ denote the length and diameter of the full ellipsoids and should not be mistaken for the length of their respective semiaxes.

Brownian particles undergo diffusive motion \cite{dhont1996introduction, HanANZLY2006}.
While translational diffusion is typically isotropic for spheres, it is anisotropic for ellipsoids. 
The diffusive motion depends on the diffusion constant, which is a scalar for spheres and a matrix for anisotropic particles.
For uniaxial particles, such as ellipsoids, there are two different translational diffusion coefficients depending on the two hydrodynamic friction coefficients $\gamma_{\mathrm{T},\parallel}$ and $\gamma_{\mathrm{T},\perp} $. The friction coefficients correspond to the friction for movement parallel and perpendicular to the particles' polar and equatorial axes, respectively~\cite{dhont1996introduction, HanANZLY2006}.
Using the Einstein-Smoluchowski equation~\cite{smoluchowski1916brownsche, vonSmoluchowski1906, Einstein1905Ueber}, these two friction coefficients lead to two diffusion coefficients. The diffusion matrix $\mymat{D}_T$ reads
\begin{align}
    \mymat{D}_T = D_{\mathrm{T},\parallel} ( \myev{u} \otimes \myev{u} ) + D_{\mathrm{T},\perp} ( \mymat{I} - \myev{u} \otimes \myev{u}), \label{eq:methods_diffusion_matrix_transl}
\end{align}
where $\mymat{I}$ is the two-dimensional identity matrix, $D_{\mathrm{T},\parallel} = k_\mathrm{B} T/\gamma_{\mathrm{T},\parallel}$ the parallel diffusion coefficient, $D_{\mathrm{T},\perp}  = k_\mathrm{B} T/\gamma_{\mathrm{T},\perp}$ the perpendicular diffusion coefficient, $\otimes$ the dyadic product, and $\myev{u}$ corresponds to the polar axis. The rotational diffusion coefficients $D_\mathrm{R}$ can be calculated using the Einstein-Smoluchowski equation~\cite{smoluchowski1916brownsche, vonSmoluchowski1906, Einstein1905Ueber}
\begin{align}
    D_\mathrm{R} = k_\mathrm{B} T / \gamma_{\mathrm{R}} 
\end{align}
where $\gamma_\mathrm{R}$ is the rotational friction coefficient of that particle.

\subsection{Lennard-Jones units and dimensionless variables}
In the simulations, we use Lennard-Jones units. Distance and energy are measured as multiples of the distance $\sigma_0$ and the energy $\epsilon$ (appearing in the interaction potential, see Appendix \ref{sec:interactionpotential}), respectively.
Friction coefficients are multiples of the friction coefficient $\gamma_{\mathrm{T},\mathrm{sp}}$ of a sphere with diameter $\sigma_0$.
Thus, the Lennard-Jones time is given by $\tlj = \sigma_0^2 \gamma_{\mathrm{T},\mathrm{sp}}/\epsilon = \sigma_0^2 \kB T/(\epsilon D_{\mathrm{T}, \mathrm{sp}})$, where $D_{\mathrm{T}, \mathrm{sp}}$ is the diffusion coefficient of a sphere that has the same volume as the ellipsoid. All parameters are measured in this unit system. For example, the active propulsion force $F_\mathrm{A}$ is given by $F_\mathrm{A} = 24 \epsilon/\sigma_0$.

As in previous work \cite{Jeggle2020,Broeker2020}, we use two dimensionless variables to parametrize the state of the system, namely the \Pe{} number for ellipsoids $ \mathrm{Pe}$ and the global packing density $\Phi_0$. The \Pe{} number is (following Ref.\ \cite{TheersWQWG2018clustering}) defined as
\begin{align}
    \mathrm{Pe} = \frac{v_0}{a_\mathrm{el} D_\mathrm{R}},
    \label{eq:Peclet_of_ellipsoid}
\end{align}
with the propulsion speed $v_0= F_\mathrm{A}D_{\mathrm{T},\parallel}/(k_\mathrm{B}T)$ of the ellipsoids, which defines $\mathrm{Pe}$ via the ratio of persistence length $v_0/D_\mathrm{R}$ and particle size $a_\mathrm{el}$. In the limit of spherical particle shapes, \cref{eq:Peclet_of_ellipsoid} corresponds to the standard definition of the \Pe{} number for active spheres, except for a factor 3. The \Pe{} number can be changed by tuning either the temperature $T$ or the propulsion force $F_\mathrm{A}$. We chose to change the \Pe{} number by changing $T$ as we would need to cope with a different effective size of the particles due to stronger propulsion when tuning the propulsion force.
Changing the temperature $T$ affects the translational and rotational diffusion and the \Pe{} number.

The global packing density is defned as $\Phi_0 = N_\mathrm{part} A_\mathrm{part}/A_\mathrm{sim}$ with the number of particles in the system $N_\mathrm{part}$, the area of a particle $A_\mathrm{part} = a_\mathrm{el} b_\mathrm{el} \pi $, and the area of the simulation domain $A_\mathrm{sim}$.

\subsection{Simulation details}
Since the influence of the particle shape on the collective dynamics has been studied thoroughly in previous work~\cite{vanDamme19, TheersWQWG2018clustering, ShiC2018self}, we focus on the impact of $\mathrm{Pe}$ and $\Phi_0$ for a fixed length-to-diameter ratio $\kappa = 2$.
The state diagram is of interest by itself, but is also required to find the parameter combinations for which the particle distribution stays homogeneous (which will later be used for obtaining the pair-distribution function). Thus, we performed computer simulations of active ellipsoids by numerically solving \cref{eq:dotr,eq:dotphi} for different \Pe{} numbers $\mathrm{Pe}$ and global packing densities $\Phi_0$. Then, we classified the observed states. 

The simulations were performed with a modified version of the molecular dynamics simulation package \lammps{}~\cite{Plimpton1995}. We modified the software package such that overdamped dynamics could be simulated. For integrating the equations of motion, we used the Euler-Maruyama method. The simulations start with random initial positions in a quadratic simulation domain with a side length of $ 128 \sigma_0$ and periodic boundary conditions. Then, the particles are simulated for $500 \tlj$ with a time step size of $2\cdot 10^{-5} \tlj$. 
\begin{figure*}
    \centering
    \includegraphics[width = \textwidth]{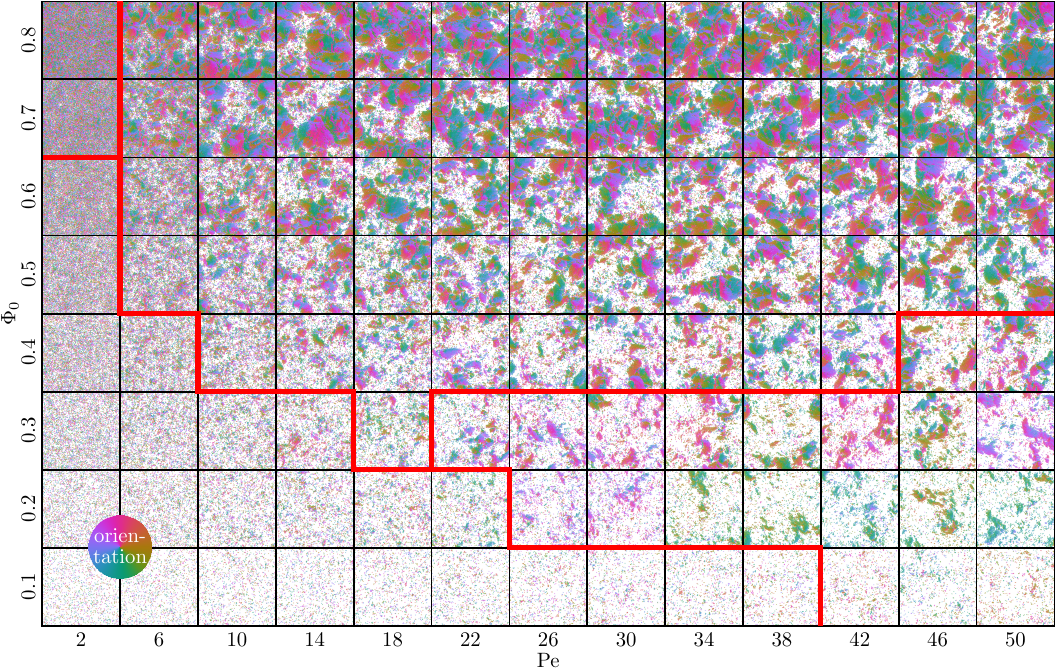}
    \caption{Snapshots of particle distributions for different \Pe{} numbers $\mathrm{Pe}$ and packing densities $\Phi_0$. Each particle is colored according to its orientation. The system is in a disordered state for low \Pe{} numbers and low densities. Local polar order emerges for high packing densities and low \Pe{} numbers. For large \Pe{} numbers, polar flocks can be seen for low packing densities and polar domains for high packing densities. The red line corresponds to the borders between these different states obtained from order parameters shown in \cref{fig:state_diagram} and visual inspection.}
    \label{fig:snaps_overview}
\end{figure*}
Note that this time step is used for all simulations. Further details regarding the computer simulations are given in Ref.\ \cite{Broeker2023Diss}.

\subsection{Classification of states}
\begin{figure*}
    \includegraphics[width=\textwidth]
    {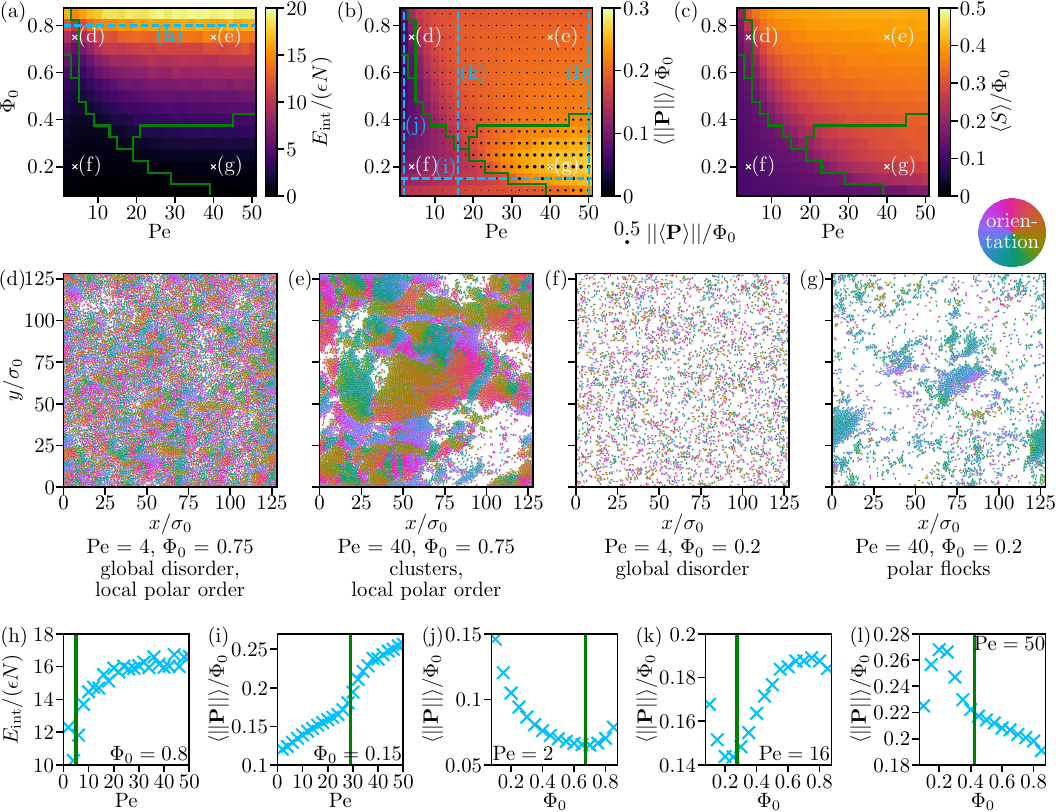}
    \caption{\textbf{Top row}: State diagrams showing (a) interaction energy per particle $E_\mathrm{int}/(\epsilon N)$, (b) local and global polarization divided by density, $\langle \norm{\myvec{P}}\rangle / \Phi_0$ and $\norm{\langle \myvec{P}\rangle} / \Phi_0$, respectively, and (c) average nematic order parameter divided by density $\langle S \rangle / \Phi_0$ as a function of \Pe{} number $\mathrm{Pe}$ and packing density $\Phi_0$. The green lines represent the state borders, determined by analyzing the order parameters and visual inspection. The cyan dashed lines represent exemplary cuts, for which the corresponding data is plotted in the bottom row in panels (h)-(l). \textbf{Middle row}: Exemplary snapshots of the different states of the system. The snapshots show
    (d) a disordered state with local polarization, (e) polarized clusters with topological defects and low-density areas, (f) a homogeneous distribution, and (g) two large polarized flocks. \textbf{Bottom row}: Example plots of the order parameters used to determine the state borders, showing (h) $E_\mathrm{int}/(\epsilon N)$ as a function of $\mathrm{Pe}$, (i) $\langle \norm{\myvec{P}}\rangle / \Phi_0$ as a function of $\mathrm{Pe}$, and (j)-(l) $\langle \norm{\myvec{P}}\rangle / \Phi_0$ as a function of $ \Phi_0$. The green lines represent the state borders. In the state diagrams in the top row, the cuts corresponding to these plots are marked by cyan lines.
    \label{fig:state_diagram}}
\end{figure*}

The particle distributions obtained in the simulations for various values of $\Phi_0$ and $\mathrm{Pe}$ are shown in \cref{fig:snaps_overview}. Each box shows an entire simulation domain. The coloring of the particles indicates their orientation. A red line is used to visualize the state borders. Note that, compared to those between MIPS and a homogeneous distribution for spherical particles, the borders between these phases are not sharp. The state diagram is shown in Figs.\ \ref{fig:state_diagram}(a)-(c) with various order parameters (interaction energy per particle in \cref{fig:state_diagram}(a), averaged local polarization divided by density in \cref{fig:state_diagram}(b), and averaged local nematic order divided by density in \cref{fig:state_diagram}(c)). Snapshots of the four states we distinguish between are shown in \cref{fig:state_diagram}(d) (global disorder with local polar order), \cref{fig:state_diagram}(e) (clusters with local polar order), \cref{fig:state_diagram}(f) (global disorder), and \cref{fig:state_diagram}(g) (polar flocks). Finally, the dependence of the order parameters on $\mathrm{Pe}$ (Figs. \ref{fig:state_diagram}(h),(i)) and $\Phi_0$ (Figs. \ref{fig:state_diagram}(j)-(l)) is used to determine the state borders (see Section \ref{sec:quant}).

We now discuss the classification of the observed states in more detail. In the regime of high densities and high \Pe{} numbers, we find polar clusters that collide with other polar clusters and thereby form larger clusters with different polar domains.
Topological defects appear at the edges of the polar domains as the polar clusters collide, and dilute areas with low densities emerge.
A larger snapshot in \cref{fig:state_diagram}(e) shows an example of these polar clusters.
Typically, these defects have a half-integer topological charge, as known for active nematics~\cite{Gromann2020, Doostmohammadi2018, PalmerCGYG2022understanding} (and nematics in general). Note that these structures are, in general, not stable as they are highly polarized, and polarization is coupled to mass transport in active matter.
Although the interaction potential (see Appendix \ref{sec:interactionpotential}) only includes nematic effects and does not explicitly include polar effects, polar order emerges. This is caused by the interplay of aligning torques and active propulsion, as has also been found experimentally in the case of the gliding bacterium \textit{Myxococcus xanthus}~\cite{PeruaniSJSDB2012collective, HarveyATA2013continuum}.   

Focusing on the density distribution at high \Pe{} numbers and packing densities and disregarding the particle orientations, one might assume that the small clusters are MIPS clusters, which are common in systems of spherical active particles. However, the clusters found here differ from MIPS clusters in two ways. First, the polar clusters are highly mobile due to their polarity and the resulting mass transport. Their movement even becomes more persistent as the orientation of the particles inside the cluster is stabilized due to the aligning interactions inside the cluster. In contrast, MIPS clusters only move via diffusion, decreasing their diffusivity with cluster size. 
The second difference is that the outer layer of MIPS clusters from spherical particles consists of particles pointing inwards and exerting pressure onto the particles inside the cluster~\cite{solon2018generalized, Caprini2MP020spontaneous}, which can also lead to a higher interaction energy~\cite{BickmannBJW2021chirality}.
On the other hand, the outer layers of particles in polar clusters are parallel to the interface between polar and nonpolar regions and, therefore, do not exert a pressure on the inner layers~\cite{Gromann2020}.

In the case of very high packing densities (such as $\Phi_0 = 0.75$), these polar regions also emerge for low \Pe{} numbers, but the system is in global disorder. There are few dilute areas, and the polar clusters are relatively small.
A snapshot of this configuration is shown in \cref{fig:state_diagram}(d). Note that topological defects can also arise in a nematic phase~\cite{BaerGHP2020rodReview, KeberLSCGBMDB2014topology, HardouinHDLLYIJS2019reconfigurable}
and become more pronounced in confinement~\cite{WittmannCLA2021particle,MonderkampWtVVWL2022}.
When comparing low and high \Pe{} numbers at high packing densities, low \Pe{} numbers correspond to high temperatures and result in a  system with global disorder and local polar order. In contrast, high \Pe{} numbers correspond to low temperatures, resulting in a more ordered system with polar clusters. %
This is similar to the observations made in Ref.~\cite{Gromann2020}, where it was found that small anisotropy leads to global disorder. Increasing the anisotropy of the particles increases the order in the system, and therefore polar clusters emerge.

If the overall packing density is low, polar clusters also emerge, but take up a substantial amount of the surrounding particles so that barely any other polar cluster exists.
This means that it is rather unlikely for such a polar cluster to collide with another polar cluster, such that the clusters can move freely.
The clusters are highly polarized, and topological defects are rare.
An example is given in \cref{fig:state_diagram}(g).
Such moving polar clusters are also observed in biology, for example in flocks of birds and schools of fish~\cite{MarchettiJRLPMS2013hydrodynamics, Vicsek1995}. Spherical ABPs with polar alignment also exhibit polar clusters~\cite{MartinLDP2018collective}.
This behavior is also known for long-range orientation ordering as in the Vicsek model~\cite{Vicsek1995} and the Toner-Tu model~\cite{TonerTu1995long, TonerT1998flocks, MahaultGC2019quantitative, TonerTR2005hydrodynamics}.

For low packing densities $\Phi_0$ and low \Pe{} numbers $\mathrm{Pe}$, the system is in a state of global disorder.
The temperature is sufficiently high to destabilize polar clusters, and the occurrence of local polar order is prevented, as in the case of high densities. An example is shown in \cref{fig:state_diagram}(f).
Small polar aggregates, which can also form in this state, are unstable.

\subsection{\label{sec:quant}Quantitative analysis of the state diagram}
For a quantitative understanding, we need to analyze which parameter combinations correspond to which state. In particular, it is important to identify the parameter combinations that lead to a homogeneous distribution of the particles, as these will later be used to compute the pair-distribution function.

Visual inspection allows for this identification only to a certain degree. For an objective distinction of the states, the interaction energy per particle, the polarization, and the nematic order for different values of the overall density and \Pe{} number need to be calculated. As these variables scale with the number of particles, we calculate the reduced interaction energy per particle $E_\mathrm{int}/(\epsilon N)$, the average local polarization divided by density $\langle \norm{\myvec{P}}\rangle/\Phi_0$, the average global polarization divided by density $\norm{ \langle  \myvec{P}\rangle}/\Phi_0$, and the average nematic order divided by density $\langle S \rangle /\Phi_0$. Here, $\norm{\cdot}$ is the Euclidean norm.
The interaction energy of the particles can be calculated straightforwardly from the interaction potential. Details regarding the definitions and calculations of the average local polarization, the average global polarization, and the average nematic order are given in Appendix \ref{sec:orderparameters}. Here, we briefly summarize their intuitive significance: The average local polarization increases if the particles form local polar clusters or polarized structures, i.e., if the particles locally have roughly the same orientation. Similarly, the averaged global polarization increases if the whole system is polarized, i.e., if all particles have roughly the same orientation.
The emergence of nematic phases, even if only local, corresponds to an increased average nematic order, i.e., to (local) alignment or anti-alignment of the particles.
We omit the terms \ZT{averaged} and \ZT{divided by density} of the order parameters from here on, i.e., we refer to the \ZT{averaged local orientation divided by density} just by \ZT{local polarization} (and similarly for other order parameters).

The interaction energy $E_\mathrm{int}$, local polarization $\langle \norm{\myvec{P}}\rangle/\Phi_0$, and nematic order $\langle S \rangle /\Phi_0$ are shown in Figs.\ \ref{fig:state_diagram}(a), (b), and (c), respectively. In Figs.\ \ref{fig:state_diagram}(b), the global polarization $ \norm{\langle  \myvec{P} \rangle }  / \Phi_0$ is shown as black dots. The area of the black dots scales linearly with the global polarization.
We extract the values of the order parameters from the simulations.
After discarding an initial $200 \tlj$ to allow for a relaxation of the system, we extracted the order parameters every $\tlj$ for the remaining $300 \tlj$ of the simulation time.  
The order parameters are then used to determine borders between the different states in addition to visual inspection of the simulation results.

The state of global disorder, observed for low \Pe{} numbers and low packing densities, features neither high interaction energies nor a measurable local or global polarization or a high nematic order as seen in Figs.\ \ref{fig:state_diagram}(a)-(c). For low \Pe{} numbers, the local polarization of dilute systems ($\Phi_0 = 0.1$) is increased compared to systems with moderate densities ($\Phi_0 = 0.3$).
As particle interactions become rare at low densities, the local polarization is not determined by multiple interacting particles but by single particles.
Therefore, the local polarization barely depends on the \Pe{} number.
However, for high \Pe{} numbers, the local and global polarization increase significantly.

The difference between the states of global disorder with and without local polar order is the increased local polarization of the former one. Typically, for low \Pe{} numbers, an increase in density reduces the average local polarization. This is shown in \cref{fig:state_diagram}(j), where the average local polarization is plotted against the density for $\mathrm{Pe} = 2$. However, at a packing density of roughly $\Phi_0 \approx0.65$, the polarization increases with the packing density.
We consider this change of the dependence of the polarization on the density to be the border between the two states of global disorder with and without local polar order.

All order parameters represent a difference between the states of global disorder and clusters with local polar order (except for the global polarization, which is small in both phases). A significant increase in interaction energy, local polarization, and nematic order corresponds to this state border.

The interaction energy is a helpful order parameter to distinguish between clusters and global disorder with local polar order.
As different polar clusters collide, the particles at the contact line are strongly pushed against each other, resulting in a high interaction energy.
Note that there is a significant difference to the increased interaction energy of spherical particles in clusters~\cite{BickmannBJW2021chirality, Stenhammar2014}: In MIPS clusters, the particles are pushed against each other due to boundary particles exerting pressure onto the particles inside, resulting in high pressure inside a MIPS cluster.
In the case of ellipsoidal particles, however, there is no additional pressure from the outside layer because the particles at the outer layer are also aligned to the polarization of a cluster. The high interaction energy stems from the collision of polar clusters.

The border between global disorder with local polar order and clusters with local polar order is determined by the interaction energy.
Both an increasing temperature (low \Pe{} numbers) and colliding polar clusters (high \Pe{} numbers) cause high interaction energies.
Hence, the \Pe{} number for which the interaction energy starts to increase determines this border (cf. \cref{fig:state_diagram}(h)).
A transition line between global disorder and clusters can be identified by considering the local polarization as well as the nematic order. As an example, we plot in \cref{fig:state_diagram}(k) the polarization as a function of $\Phi_0$. \Cref{fig:state_diagram}(k) shows that, for small packing densities $\Phi_0$, the polarization decreases with increasing $\Phi_0$. For larger densities, in contrast, the polarization grows with $\Phi_0$. We place the border between the cluster and the global disorder phase at the point where the polarization has a local minimum as a function of $\Phi_0$, which is roughly at $\Phi_0 \approx 0.275$ for $\mathrm{Pe} = 16$. This is similar to \cref{fig:state_diagram}(j). 

The state of polar flocks is characterized by a very high local polarization as well as a global polarization of the whole system. Since the global polarization only increases for polar flocks, it seems to be a convenient order parameter to characterize the system's state. However, the global polarization of a system depends on its size. The smaller the system, the easier it is to be globally polarized. Similarly, the local polarization depends on the system size if the system size is small compared to the persistence length of the particles. In our case, the system size is sufficiently large to strongly reduce this dependence. %
Therefore, we choose the local polarization to determine the border between polar flocks and other states. When approaching the state of polar flocks, either from global disorder or from polarized clusters, a sharp increase in the local polarization occurs, which is shown in Figs.\ \ref{fig:state_diagram}(i) and (l). Note that in \cref{fig:state_diagram}(i) the state of polar flocks is approached when increasing $\mathrm{Pe}$ (from left to right in the figure), whereas in \cref{fig:state_diagram}(l) it is approached for decreasing $\Phi_0$ (from right to left in the figure).

The local polarization is also high for polar clusters. However, the numerous collisions and resulting topological defects decrease the local polarization, such that it is reduced compared to the case of polar flocks. Moreover, the snapshot in \cref{fig:state_diagram}(e) shows many particles in clusters facing the opposite direction of the cluster's polarization.
Thus, these particles increase the nematic order but strongly decrease local polarization.
Particles facing the opposite direction of the clusters' polarization are extremely rare in the case of polar flocks.

\section{\label{section:3_Pair_correlation_function_ellipsoids}Pair-distribution function}
The classification of states developed in Section \ref{section:State_diagram_ellipsoids} reveals that the state of global disorder is the only one with translational and rotational symmetry. As our analysis of the pair-distribution function requires the state to possess these invariances, we restrict our attention to this state in the following.

Intuitively, the pair-distribution function $g$ measures how likely it is to find a particle at a certain position in a certain state (in our case specified by the particle orientation $\myev{u}$) given that another particle is at a certain position in a certain state. Thus, it is closely related (though not exactly identical) to a conditional probability. Formally, if $P( \{ \myvec{r}_i \}, \{ \myev{u}_i \}, t)$ is the probability for a system to be in the microscopic configuration given by the coordinates $\myvec{r}_i$ and $\myev{u}_i$ at time $t$, the $n$-particle density can be defined as 
\begin{align}
&\varrho^{(n)}(\myvec{r}_1, ..., \myvec{r}_2, \myev{u}_1, ..., \myev{u}_n, t) \\
&= \frac{N!}{(N-n)!}\Bigg( \prod_{i = n+1}^{N} \int_{\mathds{R}^2}\!\!\!\mathrm{d}^3 r_i \int_{\mathbb{S}_1} \!\!\! \mathrm{d}u_i \Bigg) P(\{ \myvec{r}_i \} , \{ \myev{u}_i \}, t)
\label{eq:rhon}
\end{align}
with the unit sphere $\mathbb{S}_1$ in two spatial dimensions. We can then further define the pair-distribution function $g$ as~\cite{HansenMD2009}
\begin{align}
    g(\myvec{r}_1, \myvec{r}_2, \myev{u}_1, \myev{u}_2, t)
    = \frac{\varrho^{(2)} ( \myvec{r}_1, \myvec{r}_2, \myev{u}_1, \myev{u}_2, t)}
    {\varrho^{(1)} (\myvec{r}_1, \myev{u}_1, t)  \varrho^{(1)} (\myvec{r}_2, \myev{u}_2, t)} \, .
    \label{eq:general_definition_of_g}
\end{align}
Given the definitions \eqref{eq:rhon} and \eqref{eq:general_definition_of_g}, $  \varrho^{(1)} (\myvec{r}_2, \myev{u}_2, t) g(\myvec{r}_1, \myvec{r}_2, \myev{u}_1, \myev{u}_2, t)/(N-1)$ is the conditional probability of finding a particle at position $\myvec{r}_2$ with orientation $\myev{u}_2$ at time $t$ given that there is another particle at the position $\myvec{r}_1$ with the orientation $\myev{u}_1$~\cite{WeberS2012} at the same time. A more detailed explanation is given in Ref.\ \cite{Broeker2020}. For the sake of simplicity, we will often (with slight abuse of terminology) refer to $g$ as a \ZT{probability} or \ZT{probability modification}.
\subsection{\label{subsection_parametrization_of_g}Parameterization of the pair-distribution function}
Next, we simplify the dependence of $g$ on the positions and orientations of the particles. In general, the pair-distribution function $g$ depends on two position and orientation vectors and on time $t$. The function $g$ will generally take a different form for different values of $\mathrm{Pe}$ and $\Phi_0$, i.e., it depends also on these parameters. In two spatial dimensions, $g$ thereby depends on $9$ independent variables.

\begin{figure}
    \centering
    \includegraphics{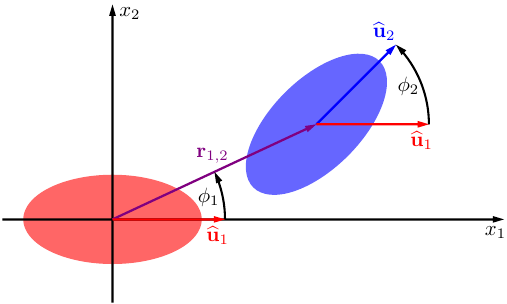}
    \caption{Coordinate system used to parameterize the pair-distribution function $g(r, \phi_1, \phi_2)$. The vector $\myvec{r}_{1,2} = \myvec{r}_2 - \myvec{r}_1$ points from the center of the first particle (red) to the center of the second particle (blue). The relative position of the second particle is determined by the length $r = \norm{\myvec{r}_{1,2}}$ of $\myvec{r}_{1,2}$ and the angle $\phi_1$ between $\myvec{r}_{1,2}$ and the orientation unit vector $\myev{u}_1$ of the first particle. The orientation of the second particle $\myev{u}_2$ is given by the angle $\phi_2$ between $\myev{u}_2$ and $\myev{u}_1$.}
    \label{fig:relative_position}
\end{figure}

To simplify the pair-distribution function, we assume that it satisfies translational and rotational invariance and that it does not depend on time. These conditions are satisfied in a homogeneous state where phase separation does not occur, which is why we numerically examine the pair-distribution function of active ellipsoids under these conditions. Given translational and rotational invariance, $g$ can only depend on the relative positions and orientations of the particles. We introduce a new coordinate system whose origin is located at the center of mass position of the first particle $\myvec{r}_1$), and align the $x_1$-axis of the new coordinate system with the orientation $\myev{u}_i$ of this particle. This coordinate system is visualized in \cref{fig:relative_position}.
The relative position of the second particle $\myvec{r}_{1,2}$ is parameterized as
\begin{equation}
\myvec{r}_2 - \myvec{r}_1 = r\myev{u}(\phi_1)    
\end{equation}
with the norm
\begin{align}
r = \norm{\myvec{r}_{1,2}} = \norm{\myvec{r}_2 - \myvec{r}_1}
\end{align}
and the angle $\phi_1$ between $\myvec{r}_{1,2}$ and the orientation of the first particle $\myev{u}_1$.
With the angle $\varphi_{1,2} $ between the center-to-center vector $\myvec{r}_{1,2}$ and the $x_1$-axis, $\phi_1$ is given by 
\begin{align}
    \phi_1 = \varphi_{1,2} - \varphi_1 \, .
\end{align}
The orientation of the second particle is parameterized by the angle $\phi_2$ between the orientation unit vectors $\myev{u}_2$ and $\myev{u}_1$, resulting in 
\begin{align}
    \phi_2 = \varphi_2- \varphi_1 \, .
\end{align}
Therefore, $g(\mathrm{Pe}, \Phi_0; r, \phi_1, \phi_2)$ depends on the three variables $r$, $\phi_1$, and $\phi_2$ and the two parameters $\mathrm{Pe}$ and $\Phi_0$. It has the symmetry property
\begin{align}
    g(\mathrm{Pe}, \Phi_0; r, \phi_1, \phi_2) = g(\mathrm{Pe}, \Phi_0; r, -\phi_1, -\phi_2). \label{eq:Symmetry_of_g}
\end{align}
\subsection{\label{section:3_explicit_value_of_g}Pair-distribution function of active Brownian ellipsoids}
Now, we investigate $g$ for fixed system parameters $\mathrm{Pe}$ and $\Phi_0$. To this end, we set the \Pe{} number to $\mathrm{Pe} = 4$ and the packing density to $\Phi_0 = 0.2$, which corresponds to the setting shown in \cref{fig:state_diagram}(f), i.e., the state of global disorder.
In this section, we omit the explicit dependence on $\mathrm{Pe}$ and $\Phi_0$ in our notation for brevity. 
To explicitly calculate $g$, we perform Brownian dynamics simulations in a quadratic simulation domain with side length $256 \sigma_0$ and a total simulation time of $550 \tlj$.
We initialize the system with random particle positions and omit the first $50\tlj$. After that, the values of the positions and orientations of the particles are extracted every $0.1 \tlj$.
To improve the statistics, we repeat this procedure $10$ times. For evaluating $g$ from the simulations, a sampling of $180$ data bins for both angles $\phi_1$ and $\phi_2$ is used.
The distance $r$ is measured with an accuracy of $0.005\sigma_0$ for all values of $r<3\sigma_0$ and with an accuracy of $0.05\sigma_0$ for $3\sigma_0 < r < 10 \sigma_0$.
See Ref.\ \cite{Broeker2023Diss} for further details on the calculation of $g$.

\begin{figure*}
    \centering
    \includegraphics[width=\textwidth] {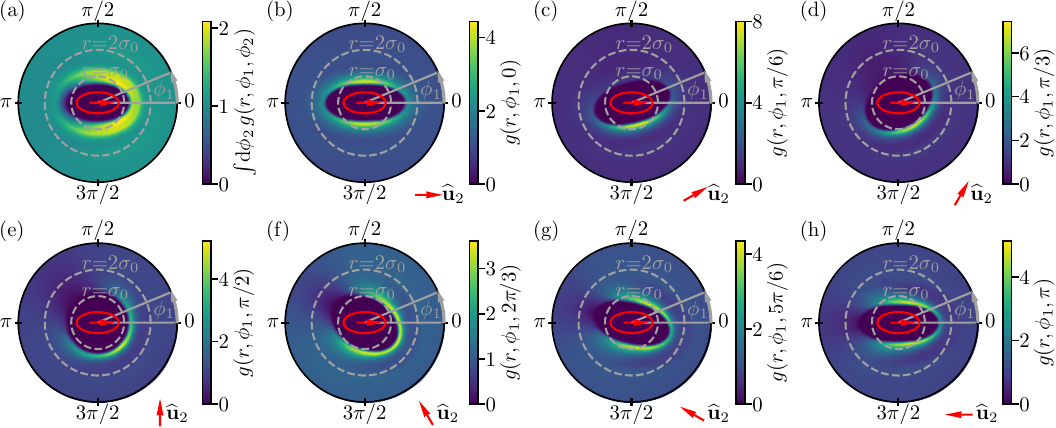}
    \caption{Pair-distribution function of active Brownian ellipsoids for the \Pe{} number $\mathrm{Pe} = 4$ and the packing density $\Phi_0 = 0.2$. (a) $g$ integrated over the orientation of the second particle. This corresponds to the probability modification of finding a second particle independent of the second particle's orientation. (b)-(h) $g(r,\phi_1, \phi_2)$ for different fixed values of $\phi_2$. The red vectors in each bottom right corner show the orientation of the second particle. The values of $\phi_2 \in (-\pi,0)$ are omitted due to the symmetry of $g$ [cf. \cref{eq:Symmetry_of_g}]. 
    \label{fig:phi2_fixed} }
\end{figure*}

The pair-distribution function $g$ depends on three parameters. Therefore, one of these parameters has to be fixed to visualize the dependence on the other two parameters.
\Cref{fig:phi2_fixed} shows $g(r, \phi_1, \phi_2)$ as a function of $r$ and $\phi_1$ for fixed values of $\phi_2$.
High values of $g$ typically result from two (not mutually exclusive) causes.
First, if a particle constellation occurs often, independent of the stability of this constellation, the corresponding value of $g$ is increased.
Second, $g$ has a high value for very stable constellations.
Both of these causes increase the probability of a constellation being measured in the simulation, and thereby the corresponding values of $g$.

Figure \ref{fig:phi2_fixed}(a) shows the integral of $g$ over the second particle's orientation ($\int\! \mathrm{d}\phi_2 \, g(r, \phi_1, \phi_2)$), which corresponds to the modification of the probability of finding a second particle with any orientation at the relative position denoted by $r$ and $\phi_1$.
In the center of the plot, at very small values of $r$, an area with the shape of an ellipsoid is visible, where the values are zero.
The interaction force of the particles allows for minor overlapping, but for strong overlapping, the interaction force becomes extremely strong, preventing two particles at the same location.
Hence, the smallest values of $r$ for which $g$ is nonzero correspond to the distance $r$ for which the particles are touching. This distance depends on $\phi_1$ since the particles are anisotropic.
At $\phi_1 \approx 0 $ and $\phi_1 \approx \pi $, the values for $r$ are roughly $r \approx 1.2 \sigma_0$ which equals the sum of the major and minor semiaxes $(a_\mathrm{el} +b_\mathrm{el})/2 \approx 1.2\sigma_0$.
For $\phi_1 \approx \pi/2$ and $\phi_2 \approx 3\pi/2$, the smallest value of $r$ with a nonzero $g$ is roughly $r \approx 0.8 \sigma_0 $ which equals $(b_\mathrm{el} +b_\mathrm{el})/2 \approx 0.8\sigma_0$. As a general rule of thumb, regarding any $\phi_1$, the smallest value of $r$ with a probability unequal to zero is typically the sum of the size of the ellipsoid at this $\phi_1$ and the minor semi axis $b_\mathrm{el}$.
The size of the reference particle is also shown in the figure.

The probability of finding another particle is roughly doubled in the proximity in front of and next to a reference particle.
This is a typical feature of ABPs~\cite{Bialk2013, SchwarzendahlM2018, Jeggle2020}.
Furthermore, the area with increased probability is relatively large compared to that of spheres~\cite{Bialk2013}.
This area consists of two maxima with a local minimum in the center. For $\phi_1 \approx 0$, one local local maximum is at  $r=(a_\mathrm{el}+b_\mathrm{el})/2$ and the other maximum is at $r=2a_\mathrm{el}/2$. There is a (barely visible) local minimum between these local maxima. Configurations where the particles touch have an increased probability, in particular if the particles are parallel or perpendicular to each other. The probability is increased by a small amount at $r \approx 2 \sigma_0$ because particles sometimes form small local structures. The distance $r \approx 2 \sigma_0$ corresponds to the next but one particle.
This is again very similar to the case of spheres~\cite{Bialk__2015}, but the effect is weaker as the maximum is less pronounced due to the anisotropy of the shape.

In Figs.\ \ref{fig:phi2_fixed}(b)-(h), the pair-distribution function $g$ is shown for different fixed angles of the second particle's orientation $\phi_2$. Figure \ref{fig:phi2_fixed}(b) shows $g(r, \phi_1, 0)$, i.e., the case where the second particle is parallel to the reference particle. The probability is strongly increased for particles close to the reference particle, i.e., angles $\phi_1 = \pi/2$ and $\phi_1 = 3\pi/2$.
Note also that the area in the center, where the values of $g$ are zero, adapts the particle shape. As we only consider parallel particles in Fig.\ \ref{fig:phi2_fixed}(b), the values of $r$ at which the particles would severely overlap strongly depend on $\phi_1$.
At $\phi_1 \approx 0 $ and $\phi_1 \approx \pi$, the lowest value of $r$ for which another particle is found is $r \approx 2 a_\mathrm{el}/2$, i.e., twice the major semi axis, and for $\phi_1 \approx \pi/2 $ and $\phi_1 \approx 3\pi/2 $, it is $r \approx 2 b_\mathrm{el}/2$, i.e., twice the minor semi axis.
For Figs.\ \ref{fig:phi2_fixed}(b)-(h), the shape of the area with $g=0$ depends on the second particle's orientation as strong overlapping is not possible.

Suppose that we fix the reference particle in the center and move the other particle around the reference particle with a stationary orientation while maintaining contact between the particles. In that case, the resulting path of the center of the second particle determines the shape of the inner low probability zone. For parallel particles, this results in a thin ellipse (cf. Figs.\ \ref{fig:phi2_fixed}(b) and (h)), and for perpendicular particles, the resulting area is nearly a perfect circle as shown in \cref{fig:phi2_fixed}(e). Other orientations lead to a shape similar to a rotated ellipse. 
In cases where the particles are roughly pointing in the same direction, like the ones shown in \cref{fig:phi2_fixed}(c) ($\phi_2=\pi/6$) and (d) ($\phi_2=\pi/3$), the highest probability peaks are located where the particles push into each other's paths. Both particles swim against each other, resulting in a relatively stable position.
In the case of perpendicular particles as in \cref{fig:phi2_fixed}(e) ($\phi_2=\pi/2$), the highest probability peaks also occur where the particles collide.
If the particles point roughly in opposite directions, the maximum is instead found for a constellation resulting from a collision. For example, in \cref{fig:phi2_fixed}(g) ($\phi_2=5\pi/6$) the maximum values of $g$ correspond to particles passing each other at a close distance. A similar effect can be observed in \cref{fig:phi2_fixed}(h) and is also known for spheres~\cite{Jeggle2020}. Particle constellations resulting from a previous collision have an enhanced probability of occurring.
Furthermore, particles pointing in opposing directions tend to form a probability shadow (a region of locally decreased probability). This results from the fact that particle constellations that can only arise if particles have moved through each other (which is not possible for the interaction potential used here) are very unlikely. An example is shown in \cref{fig:phi2_fixed}(e), where the values of $g$ are low for $\phi_1 \approx 3\pi/4$.
This probability shadow has also been observed for spherical ABPs~\cite{DhontPB2021motility, Jeggle2020, Bialk2013, SchwarzendahlM2018}.
The depletion zone behind an ABP arises for particles with opposite orientations as in Figs.\ \ref{fig:phi2_fixed}(g) and (h). In contrast, this depletion zone barely exists if the particles are parallel as in \cref{fig:phi2_fixed}(b) and (c). Here, the probability of finding a particle directly behind another one is increased.

Another way to show the different characteristics of $g$ is to fix the distance parameter $r$ and then display the dependence of $g$ on $\phi_1$ and $\phi_2$, as done in Refs.~\cite{Jeggle2020, Broeker2020} for spherical particles.
Such a plot can help to answer two questions:

(i): For a certain center-to-center distance, what are the typical angle constellations corresponding to high and low values of $g$?\\
(ii): How complex is the structure of $g$ with respect to the dependence on the angles?

\begin{figure*}
    \centering
    \includegraphics[width=\textwidth] {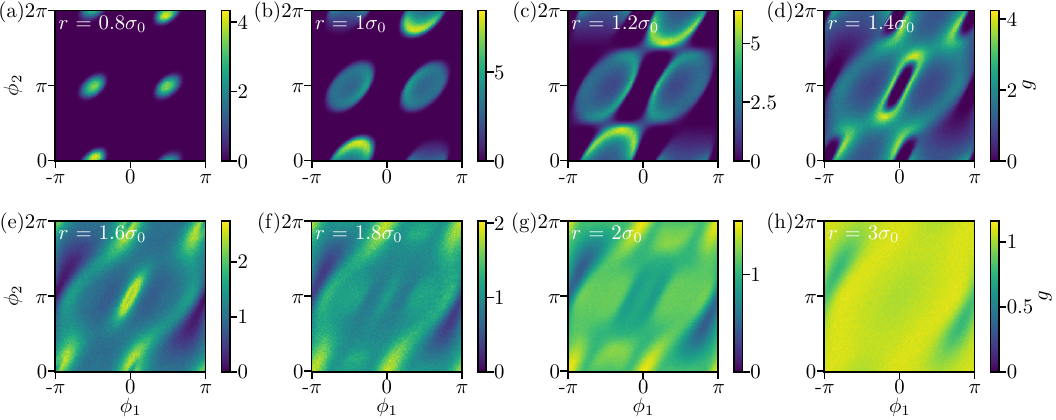}
    \caption{Pair-distribution function of active Brownian ellipsoids for fixed \Pe{} number $\mathrm{Pe} = 4$ and packing density $\Phi_0 = 0.2$ and for different values of the center-to-center distance $r$. This corresponds to the probability modification of finding a second particle at a certain distance $r$ depending on the position angle $\phi_1$ and the orientation angle $\phi_2$. The plots show a point symmetry with respect to $\phi_1 = 0$ and $\phi_2 = \pi$ due to the symmetry of $g$ of [cf.~\cref{eq:Symmetry_of_g}]. \label{fig:r_fixed} }
\end{figure*}

The latter question is essential when looking for an analytical approximation for $g$, which will be considered later using the Fourier approximation model for the dependence on the angles.
The pair-distribution function $g$ for different center-to-center distances $r$ is shown in \cref{fig:r_fixed}.
In the case of minimal distances $r = 0.8\sigma_0 $ (cf.~\cref{fig:r_fixed}(a)), particles are only found if both particles are parallel or anti-parallel and the second particle is next to the reference particle. 
In other words, their minor semi axes create a straight line. The parallel case ($\phi_2 \approx 0$ and $\phi_2 \approx 2\pi$) has a higher probability than the anti-parallel case as it is more stable.
At slightly higher distances $r=\sigma_0$, $r = 1.2 \sigma_0$, and $r = 1.4 \sigma_0$, as shown in Figs.\ \ref{fig:r_fixed}(b), (c), and (d), similar angular constellations correspond to local maxima of $g$. The maxima become broader, and a local minimum emerges at their centers. At higher distances, a larger range of angles is possible without the particles overlapping too strongly.
Additionally, the difference between the parallel and anti-parallel cases becomes more pronounced.
The maxima of $g$ are not symmetric. Instead, higher values are found if the absolute value of $\phi_1 $ is scarcely smaller than $\pi/2$, which means that the second particle is minimally further ahead than the reference particle. The values of $g$ are also increased if the second particle points towards the first particle's path ahead, corresponding to values of $\phi_2 \approx \pi/4$ for $\phi_1 = -\pi/4$.
The observation that local maxima tend to widen up, and local minima emerge in between for increasing distances has also been made for spherical particles in two~\cite{Jeggle2020} and three~\cite{Broeker2020} dimensions. At distances close to $a_\mathrm{el} \approx 1.6 \sigma_0 $ as shown in \cref{fig:r_fixed}(e), a local maximum is found for two particles swimming straight into each other with $\phi_1 = 0$ and $\phi_2 = \pi$ and a local minimum emerges for $\phi_1 = \pi$ and $\phi_2 = \pi$. The latter corresponds to a constellation that would require the particles to move through each other.
If both particles have the same orientation ($\phi_2 = 0$), local maxima can be found for $\phi_1=0$ and $\phi_1 = \pi$. If the particles are parallel, the constellation where both particles swim right behind each other has an increased probability.
Also, as $r = 1.6 \sigma_0 $ is slightly larger than $a_\mathrm{el} = 4^{1/3}\, \sigma_0$, a very thin local minimum emerges inside the maximum.
At high distances such as $r=2\sigma_0 $ and $r=3\sigma_0$ shown in Figs.\ \ref{fig:r_fixed}(g) and (h), respectively, the angular dependence of $g$ slowly vanishes. Even for $r=3\sigma_0$, the probability of the constellation of $\phi_1 = \pi$ and $\phi_2 = \pi$, which results from two particles passing through each other, is significantly decreased.
The general asymmetry of the maxima and minima distribution is similar to the spherical case~\cite{Jeggle2020, Broeker2020} and a result of the co-dependence of the angles.
Stable constellations correspond to high values of $g$, but a constellation with an offset in the position angle $ \phi_1 $ is not particularly stable. However, if the orientation of the second particle compensates for the offset, the constellation can be stable. Therefore, the maxima are distorted.
The same explanation can be used for the distortion of the minima in \cref{fig:r_fixed}(h). The minima correspond to constellations of two particles that can arise only if they pass through each other. If the second particle's position is not exactly at $\phi_1 = \pi$, but slightly different (say, $\phi_1 = \pi + \alpha$ with a small angle $\alpha$), the minimum is located at a slightly different orientation angle 
$\phi_2 \approx \pi + 2\alpha $. The additional factor of $2$ stems from the fact that the reference particle is also moving.
If the reference particle was not moving, the setting $\phi_1 = \pi + \alpha$ and $\phi_2 = \pi + \alpha$ would correspond to the second particle having moved through the reference particle.
However, as the reference particle is also active, the angles must be adjusted accordingly.
\subsection{\label{subsection:3_analytical_approximation}Analytical approximation of the pair-distribution function}
The pair-distribution function often appears in microscopic field-theoretical models for active matter~\cite{HansenMD2009, BickmannW2020, BickmannW2020b, BickmannBJW2021chirality, BickmannBroekertVW2022external, broeker2022orientation, Jeggle2020}, and an analytical expression for this function is needed in order to derive them in a closed form \cite{teVrugtBW2022}. Therefore, we here determine an analytical expression for $g$ that is valid for a wide range of \Pe{} numbers and packing densities.
For this purpose, we first use the $2\pi$-periodicity of $g(\mathrm{Pe}, \Phi_0; r, \phi_1, \phi_2)$ regarding the angles $\phi_1$ and $\phi_2$ and perform the real Fourier expansion

\begin{equation}
\begin{split}
    &g(\mathrm{Pe}, \Phi_0; r, \phi_1, \phi_2)\\ &=
     \sum_{h,j=0}^{\infty} \sum_{k,l=1}^{2} a_{h,j}^{k,l}(\mathrm{Pe}, \Phi_0; r)
    w_k(h\phi_1) w_l(j\phi_2)    
\end{split}
\end{equation}
with the functions 
\begin{align}
    w_1(x) &= \cos(x), \\
    w_2(x) &= \sin(x)
\end{align}
and the real Fourier coefficients $a_{h,j}^{k,l}(\mathrm{Pe}, \Phi_0; r)$ which depend on the system parameters $\mathrm{Pe}$ and $\Phi_0$ and the distance $r$.
The Fourier coefficients $a_{h,j}^{k,l}(\mathrm{Pe}, \Phi_0; r)$ are given by
\begin{equation}
\begin{split}
    a_{h,j}^{k,l}(\mathrm{Pe}, \Phi_0; r) &= \frac{1}{\pi^2}  \int_0^{2\pi} \!\!\!\!\! \dif \phi_1 
    \int_0^{2\pi} \!\!\!\!\!\dif \phi_2 \, g(\mathrm{Pe}, \Phi_0; r, \phi_1, \phi_2) \, \\
    &\quad w_k(h\phi_1) \, w_l(j\phi_2) \, \frac{1}{2^{\delta_{h,0} + \delta_{j,0}}}
\end{split}
\end{equation}
with the Kronecker delta $\delta_{ij}$. Further details on the Fourier calculation of discrete data of a histogram are given in Ref.\ \cite{Broeker2023Diss}.
As the pair-distribution function $g$ has the symmetry property
\begin{align}
    g(\mathrm{Pe}, \Phi_0; r, \phi_1, \phi_2) = g(\mathrm{Pe}, \Phi_0; r, -\phi_1, -\phi_2) \, ,
\end{align}
the resulting Fourier representation has the same symmetry, such that the coefficients $a_{h,j}^{1,2} = a_{h,j}^{2,1}$, which correspond to a product of a $\sin$ and a $\cos$ function in the Fourier expansion, vanish:
\begin{align}
    a_{h,j}^{1,2} = a_{h,j}^{2,1} = 0  \, \, \, \, \forall \, \,  h,j \, .
\end{align}
The coefficients that represent either two $\sin$ functions or two $\cos$ functions remain.
This leaves $g$ as
\begin{equation}
\begin{aligned}
    &g(\mathrm{Pe}, \Phi_0; r, \phi_1, \phi_2)  \\
    &= \sum_{h,j=0}^{\infty} \sum_{k=1}^{2} a_{h,j}^{k,k}(\mathrm{Pe}, \Phi_0; r)
    w_k(h\phi_1) w_k(j\phi_2) \, .
\end{aligned}
\end{equation}
In the cases of spheres~\cite{Jeggle2020, Broeker2020}, the angular dependence of $g$ can be represented reasonably accurately by a Fourier expansion truncated at second order.
In the case of ellipsoids, the angular dependence is more complex, and we use the Fourier modes up to third order:
\begin{equation}
\begin{aligned}
    &g(\mathrm{Pe}, \Phi_0; r, \phi_1, \phi_2) \approx \\
    & \sum_{h,j=0}^{3} \sum_{k=1}^{2} a_{h,j}^{k,k}(\mathrm{Pe}, \Phi_0; r)
    w_k(h\phi_1) w_k(j\phi_2) \, .
\end{aligned}
\label{eq:fourierexpansionofg}
\end{equation}
As the zero-frequency contribution of the $\sin$ functions vanishes, we find
\begin{align}
  a_{0,j}^{2,2} = 0 \, \, \, \, \forall \, \, j\, , \\
  a_{k,0}^{2,2} = 0 \, \, \, \, \forall \, \, k\, . 
\end{align}
We can therefore represent the approximation as 
\begin{equation} 
\begin{aligned}
    &g(\mathrm{Pe}, \Phi_0; r, \phi_1, \phi_2) \approx \\
    &\sum_{h,j=0}^{3} a_{h,j}^{1,1}(\mathrm{Pe}, \Phi_0; r)
    \cos(h\phi_1) \cos(j\phi_2) \\
   &+ \sum_{h,j=1}^{3} a_{h,j}^{2,2}(\mathrm{Pe}, \Phi_0; r)
    \sin(h\phi_1) \sin(j\phi_2)  \, 
\label{eq:approx_of_g}
\end{aligned}
\end{equation}
with a total of $16 + 9 = 25$ Fourier coefficients.
To obtain a full analytical representation of $g$ that is not only valid for specific values of $r$, $\mathrm{Pe}$, and $\Phi_0$, we examine the dependence of every Fourier coefficient on $r$ for all combinations of $\mathrm{Pe}$ and $\Phi_0$ and fit it via suitable functions. 

How complex the dependence of $g$ on $r$ is depends strongly on the parameters $\mathrm{Pe}$ and $\Phi_0$. Since we wish to choose the same functional form for all values of these parameters, choosing a functional form that is a good representation of $g$ for parameter values where $g$ is a complicated function may lead to overfitting in parameter regions where $g$ is simpler. To avoid problems related to overfitting, we here choose a function that is adapted to the shape that $g$ has in regions where it is a simple function of $r$.
As a particle at $\myvec{r}_1$ does not affect particles at a position $\myvec{r}_2$ far away ($\norm{\myvec{r}_2 - \myvec{r}_1 } \gg \sigma_0$), the pair-distribution function $g$ converges to $1$ for very large distances.
Thus, all Fourier coefficients must vanish for large $r$ except for $a_{0,0}^{1,1}$, which must converge to one.
All Fourier coefficients must converge to zero for small values of $r$ since two particles cannot be at the same location.

Many Fourier coefficients can be well represented by either a product of the exponentially modified Gaussian distribution function (EMG function) and a polynomial or the product of a Gaussian function and a polynomial.
The EMG function is given by
\begin{align}
\begin{split}
  \mathrm{EMG}(r; \mu, \stddev, \lambda) &= \frac{\lambda}{2} \exp \Big( \frac{\lambda}{2} ( \lambda \stddev^2  -2(r-\mu)) \Big)  \\
&\quad \: \:  \mathrm{erfc}\Big(\frac{ \lambda \stddev^2 - (r-\mu)}{\sqrt{2} \stddev}\Big), \label{Eq:ch3_Emg_function}
\end{split} 
\end{align}
where $\mu$, $\stddev$, and $\lambda$ are the mean value, standard deviation, and control parameter of the skewness of the distribution, respectively, and $\mathrm{erfc}$ is the complementary error function
\begin{align}
    \mathrm{erfc}(x) = \frac{2}{\pi} \int_x^\infty \!\!\!\!\! \dif t \exp({-t^2})  \, .
\end{align}
The EMG function is similar to the Gaussian function 
\begin{align}
    f_\mathrm{G}(r; \mu, \stddev) = \frac{1}{\stddev \sqrt{2\pi}} \exp \Big( \frac{-(r - \mu)^2}{2 \stddev^2} \Big),
\end{align}
where the parameter $\lambda$ is used to skew the function. For the coefficient $a_{0,0}^{1,1}$, which is supposed to converge to one for very large $r$, a sum of the EMG function and the tangens hyperbolicus 
\begin{align}
 \tanh(x)  =  \frac{\exp(x) - \exp(-x)}{\exp(x) + \exp(-x)}
\end{align}
is used. The argument of the $\tanh$ function is shifted to the center of the EMG function and multiplied by an additional fit parameter that controls its steepness. Also, the $\tanh$ is shifted and scaled to the range of $[0,1]$.
The fit functions for the other Fourier coefficients are chosen on an empirical basis by investigating their behavior for all values of $\mathrm{Pe}$ and $\Phi_0$.
The most suitable functions for each Fourier coefficient are shown in \cref{table:g_fit_functions}. We use these functions to fit every Fourier coefficient for every combination of $\mathrm{Pe}$ and $\Phi_0$ that corresponds to the state of global disorder as shown in \cref{fig:state_diagram}.
\begin{table}
\centering
\begin{tabular}{|c | c|} 
 \hline
 Fourier coefficient & fit function \\ 
 \hline\hline
$a_{0,0}^{1,1}$ & $a\,\mathrm{EMG}(r; \mu, \stddev, \lambda) + \tanh( (r-\mu)l_1)/2 + 0.5 $  \\  
$a_{0,1}^{1,1}$ & $a_1\,f_\mathrm{Gauss}(r; \mu_1, \stddev_1) + a_2\,f_\mathrm{Gauss}(r; \mu_2, \stddev_2)$ \\
$a_{0,2}^{1,1}$ & $a\,\mathrm{EMG}(r; \mu, \stddev, \lambda) (r^2 + l_1 r + l_2)$ \\
$a_{0,3}^{1,1}$ & $a\,f_\mathrm{Gauss}(r; \mu, \stddev)$ \\
$a_{1,0}^{1,1}$ & $a\,\mathrm{EMG}(r; \mu, \stddev, \lambda) (r^2 + l_1 r + l_2)$ \\  
$a_{1,1}^{1,1}$ & $a\,\mathrm{EMG}(r; \mu, \stddev, \lambda) (r^2 + l_1 r + l_2)(r-l_3)$ \\
$a_{1,2}^{1,1}$ & $a\,\mathrm{EMG}(r; \mu, \stddev, \lambda) (r^2 + l_1 r + l_2)(r-l_3)$ \\
$a_{1,3}^{1,1}$ & $a\,f_\mathrm{Gauss}(r; \mu, \stddev)$ \\
$a_{2,0}^{1,1}$ & $a\,\mathrm{EMG}(r; \mu, \stddev, \lambda) (r-l_1)(r-l_2)$ \\  
$a_{2,1}^{1,1}$ & $a\,\mathrm{EMG}(r; \mu, \stddev, \lambda) (r-l_1)$ \\
$a_{2,2}^{1,1}$ & $a\,\mathrm{EMG}(r; \mu, \stddev, \lambda) (r-l_1)$ \\
$a_{2,3}^{1,1}$ & $a\,f_\mathrm{Gauss}(r; \mu, \stddev)$ \\
$a_{3,0}^{1,1}$ & $a\,\mathrm{EMG}(r; \mu, \stddev, \lambda) (r-l_1)(r-l_2)(r-l_3)$ \\  
$a_{3,1}^{1,1}$ & $a\,f_\mathrm{Gauss}(r; \mu, \stddev)(r-l_1)(r-l_2)$ \\
$a_{3,2}^{1,1}$ & $a\,\mathrm{EMG}(r; \mu, \stddev, \lambda) (r-l_1)$ \\
$a_{3,3}^{1,1}$ & $a\,\mathrm{EMG}(r; \mu, \stddev, \lambda) (r-l_1)(r-l_2)$ \\
\hline
$a_{1,1}^{2,2}$ & $a\,\mathrm{EMG}(r; \mu, \stddev, \lambda) (r^2 + l_1 r + l_2)$ \\
$a_{1,2}^{2,2}$ & $a\,\mathrm{EMG}(r; \mu, \stddev, \lambda) (r^2 + l_1 r + l_2)$ \\
$a_{1,3}^{2,2}$ & $a\,\mathrm{EMG}(r; \mu, \stddev, \lambda) (r-l_1)(r-l_2)$ \\
$a_{2,1}^{2,2}$ & $a\,\mathrm{EMG}(r; \mu, \stddev, \lambda) (r-l_1)$ \\
$a_{2,2}^{2,2}$ & $a\,\mathrm{EMG}(r; \mu, \stddev, \lambda) (r-l_1)(r-l_2)$ \\
$a_{2,3}^{2,2}$ & $a\,f_\mathrm{Gauss}(r; \mu, \stddev)(r-l_1)(r-l_2)$ \\
$a_{3,1}^{2,2}$ & $a\,f_\mathrm{Gauss}(r; \mu, \stddev)(r-l_1)$ \\
$a_{3,2}^{2,2}$ & $a\,\mathrm{EMG}(r; \mu, \stddev, \lambda) (r-l_1)(r-l_2)$ \\
$a_{3,3}^{2,2}$ & $a\,\mathrm{EMG}(r; \mu, \stddev, \lambda) (r-l_1)(r-l_2)$ \\
[1ex] 
 \hline
\end{tabular}
\caption{Table of the fit functions used for the Fourier coefficients of the pair-distribution function appearing in \cref{eq:fourierexpansionofg}.}
\label{table:g_fit_functions}
\end{table}

Due to its additional fitting parameter, the EMG function usually reproduces a single function better than the Gaussian distribution. Still, as the courses of the Fourier coefficients differ strongly in some cases, the fitting procedure of the EMG function is unstable for some system parameters. In these cases, the Gaussian distribution is chosen.

\begin{figure*}
    \centering
    \includegraphics[width = \textwidth]{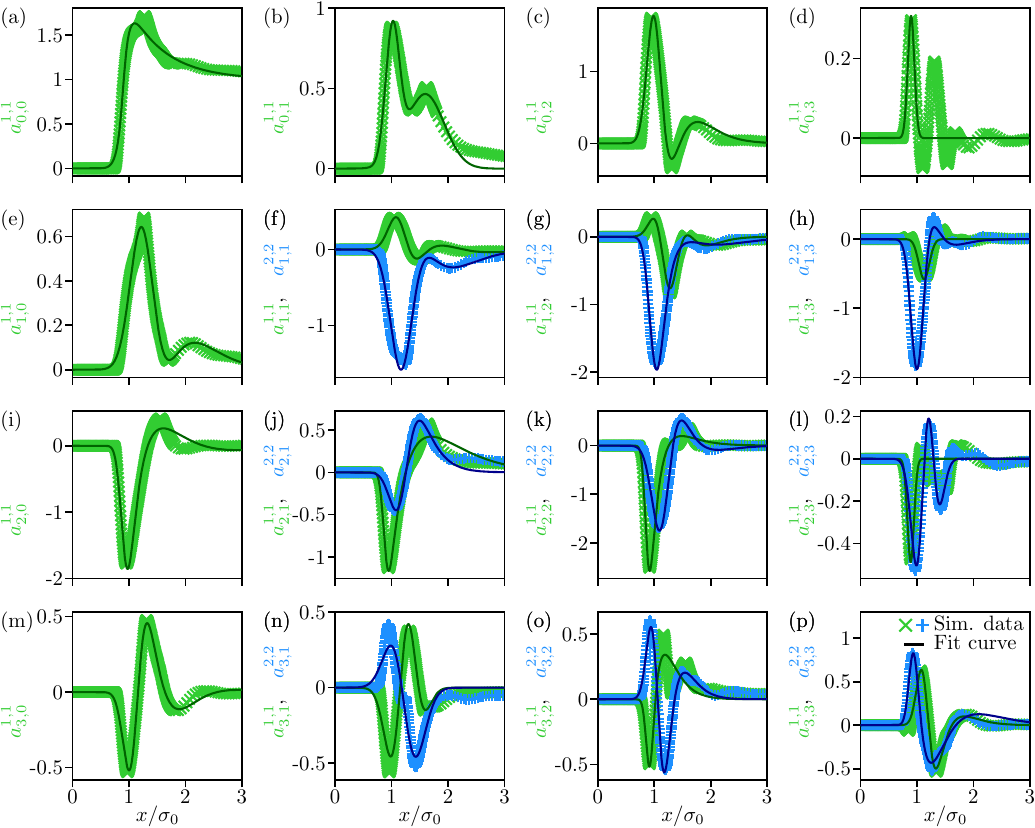}
    \caption{Fourier coefficients $a_{h,j}^{k,k}$ of $g$ for the simulation data (symbols) and the corresponding fitting results (solid lines) with the fit functions specified in \cref{table:g_fit_functions}. The coefficients $a_{h,j}^{1,1}$ correspond to green crosses and the coefficients $a_{h,j}^{2,2}$ to blue plus symbols. The row and column correspond to the $h$ and $j$ index, respectively.
    \label{fig:fourier_fit_vs_r}}
\end{figure*}

The Fourier coefficients extracted from simulation data and the fitting results for the \Pe{} number $\mathrm{Pe}= 10$ and packing density $\Phi_0 = 0.2$ are shown in \cref{fig:fourier_fit_vs_r}.
Although we calculate $g$ for the range $0 < r < 10 \sigma_0$ and fitted this entire interval, in \cref{fig:fourier_fit_vs_r} we focus on $0 < r < 3 \sigma_0$ as the Fourier coefficients show the most interesting behavior in this range.
This fitting procedure leaves us with fit parameters for each combination of $\mathrm{Pe}$ and $\Phi_0$. Hence, we offer an analytical representation of $g$ for a discrete set of system parameters.
As a last step, each fit parameter is interpolated between the values of $\mathrm{Pe}$ and $\Phi_0$ for which the fitting was performed.

Therefore, we need to ensure that the parameters of the former fitting functions are changing sufficiently smoothly regarding $\mathrm{Pe}$ and $\Phi_0$.
Choosing suitable fitting functions is a crucial step for this.
However, as some of the functions feature a lot of fitting parameters and are pretty complex, the fitting procedure improves when providing suitable starting parameters.
This is accomplished by setting the starting parameters either to expected values or values resulting from the fitting procedure for comparable \Pe{} numbers or packing densities.

The interpolation is performed by using the function
\begin{align}
h(\mathrm{Pe}, \Phi_0) = \sum_{m=-2}^{2} \sum_{n=0}^{3} \mathrm{Pe}^{\frac{m}{2}} \Phi_0^n \hparamvalue_{m,n} \, ,
\label{eq:fitfunc_PePhi}
\end{align}
with the fit parameters $\hparamvalue_{m,n}$, which is employed for every fit parameter of every Fourier coefficient.
This is again an empirical ansatz.
Note that negative exponents of $\mathrm{Pe}$ correspond to positive exponents regarding the dependence on the temperature $T$ as $\mathrm{Pe} \propto  1/T$.
Also, the course of some of the fit parameters is less complex than what \cref{eq:fitfunc_PePhi} is capable of reproducing, but for simplicity, we choose one function for all parameters.
The resulting values for $\hparamvalue_{m,n}$ for each fit parameter are shown in \crefrange{tab:param_table_1}{tab:param_table_13}.
With this interpolation, we obtain a fully analytical approximation $g_\mathrm{approx}$ of the pair-distribution function $g(\mathrm{Pe}, \Phi_0; r, \phi_1, \phi_2)$.

\begin{figure*}
    \centering
    \includegraphics[width = \textwidth]{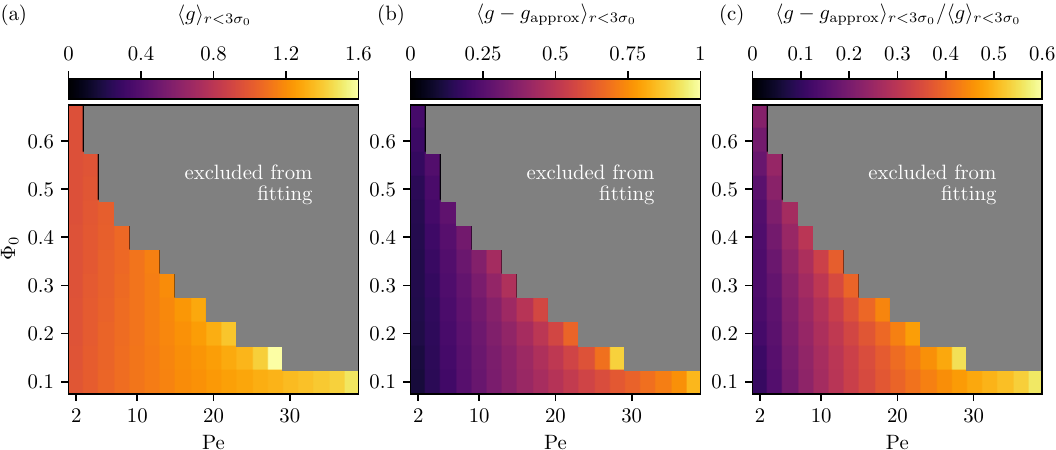}
    \caption{(a) Mean value $\langle g \rangle_{r<3\sigma_0}$ of $g$. (b)  Absolute error, i.e., difference $\langle g - g_\mathrm{approx} \rangle_{r<3\sigma_0}$ between the approximation $g_\mathrm{approx}$ and $g$. (c) Mean relative error $\langle g - g_\mathrm{approx} \rangle_{r<3\sigma_0}/\langle g \rangle_{r<3\sigma_0}$. All results shown here are calculated in the range $0<r<3\sigma_0$.
    \label{fig:error_plot}}
\end{figure*}

To summarize: To reproduce $g_\mathrm{approx}(\mathrm{Pe}, \Phi_0; r, \phi_1, \phi_2)$ for specific values of $\mathrm{Pe}$ and $\Phi_0$, the first step is to calculate each fit parameter using the function $h(\mathrm{Pe}, \Phi_0)$ in \cref{eq:fitfunc_PePhi} and the parameters given in \crefrange{tab:param_table_1}{tab:param_table_13} (see Appendix \ref{sec:tables}).
The resulting parameters are then employed in the functions in \cref{table:g_fit_functions} to create the Fourier coefficients $a_{h,j}^{k,k}$. These Fourier coefficients can be put into \cref{eq:approx_of_g} which gives $g_\mathrm{approx}(\mathrm{Pe}, \Phi_0; r, \phi_1, \phi_2)$.
To determine the quality of the analytical approximation $g_\mathrm{approx}$ for all considered values of $\mathrm{Pe}$ and $\Phi_0$, we calculate the mean absolute error $\langle g - g_\mathrm{approx} \rangle_{r<3\sigma_0}$ via
\begin{align}
    \langle g - g_\mathrm{approx} \rangle_{r<3\sigma_0} =
    \frac{
    \int_0^{3\sigma_0} \!\! \dif r
    \int_0^{2\pi} \!\! \dif \phi_1
    \int_0^{2\pi} \!\! \dif \phi_2
    |g - g_\mathrm{approx} |
    }{
     \int_0^{3\sigma_0} \!\! \dif r
    \int_0^{2\pi} \!\! \dif \phi_1
    \int_0^{2\pi} \!\! \dif \phi_2 \, 1 } \, .
\end{align}
We integrate both angles $\phi_1$ and $\phi_2$ over the full range $[0, 2\pi [$ and integrate the distance $r$ from $0$ to $3\sigma_0$.
The upper limit $3\sigma_0$ for the distance $r$ is chosen according to the typical course of $g$.
The angular dependence weakens for large $r$, which is shown in \cref{fig:phi2_fixed,fig:r_fixed} as well as in the corresponding Fourier coefficients in \cref{fig:fourier_fit_vs_r}.
Thus, $g$ is harder to reproduce for small values of $r$ and easier to reproduce for large values of $r$.
We are interested in the errors for the hard-to-reproduce range, so we chose $r<3\sigma_0$.
As the absolute error can be hard to interpret, we calculate the mean values 
$\langle g \rangle_{r<3\sigma_0}$ via 
\begin{align}
    \langle g \rangle_{r<3\sigma_0} =
    \frac{
    \int_0^{3\sigma_0} \!\!\! \dif r
    \int_0^{2\pi} \!\!\! \dif \phi_1
    \int_0^{2\pi} \!\!\! \dif \phi_2\,
    g
        }{
        \int_0^{3\sigma_0} \!\!\! \dif r
    \int_0^{2\pi} \!\!\! \dif \phi_1
    \int_0^{2\pi} \!\!\! \dif \phi_2 \, 1 } \, ,
\end{align}
allowing us to calculate the relative error $\langle g - g_\mathrm{approx} \rangle_{r<3\sigma_0}/\langle g \rangle_{r<3\sigma_0}$. The results are shown in \cref{fig:error_plot}.
In \cref{fig:error_plot}(a), the mean values of $g$, $\langle g \rangle_{r<3\sigma_0}$, are shown. The values slowly increase with an increase of the \Pe{} number and are mostly unaffected by changes in density.
As $g$ is a \ZT{probability modifcation}, the average values of $g$ can be interpreted as the modification of the probability of finding a second particle in the proximity of the reference particle. If particles tend to aggregate, the average value of $g$ increases.
The absolute error $\langle g - g_\mathrm{approx} \rangle_{r<3\sigma_0}$ is shown in \cref{fig:error_plot}(b).
Compared to the average value of $g$, it heavily depends on the \Pe{} number and grows strongly.
The absolute error slightly increases with increasing packing density. 
Figure \ref{fig:error_plot}(c) shows the relative error $\langle g - g_\mathrm{approx} \rangle_{r<3\sigma_0}/\langle g \rangle_{r<3\sigma_0}$ of the approximation. It varies between $10\%$ and $56\%$ and is especially high for high \Pe{} numbers, which correspond to low temperatures.

Generally, when reproducing the angular dependence of $g$ with Fourier modes, sharp peaks of $g$ are harder to reproduce.
If the temperature in the system is very high, the particles' interactions become more randomized, and sharp peaks of $g$ are more \ZT{washed out}. This explains the improved approximation results for low \Pe{} numbers corresponding to high temperatures.
Similar dependencies of the average value, absolute error, and relative error on the system parameters are found in the case of spherical particles~\cite{Jeggle2020, Broeker2020}.
\section{\label{sec:conclusion}Conclusion}
In this article, we have studied the collective dynamics of active Brownian ellipsoids via computer simulations and obtained their state diagram. Depending on the \Pe{} number and the packing density, the system exhibits cluster formation, local polar order, polar flocks, and disordered phases. In addition, we have provided a detailed discussion of the pair-distribution function of active ellipsoids and obtained an analytical representation of this pair-distribution function. 

Given that pair-distribution functions obtained in previous work \cite{Jeggle2020,Broeker2020} have been used as an input in field-theoretical models for active spheres \cite{BickmannW2020,BickmannW2020b}, a natural continuation of the present work would be the development of a field theory for active ellipsoids based on the pair-distribution function obtained here. Moreover, one could investigate the state diagram of active ellipsoids in three spatial dimensions in order to analyze the effects of dimensionality, which are likely to be more pronounced than for spheres. 

\section*{Supplementary Material}
The Supplementary Material \cite{SI} contains a spreadsheet with the values of the fit coefficients (as shown in Appendix \ref{sec:tables}) that are needed to recreate the analytical representation of the pair-distribution function, a Python script \texttt{abp.ellipsoidal2d.pairdistribution} that recreates the approximation of the pair-distribution function $g$ using the values of the fit coefficients, and the Python scripts and raw data needed to recreate Figs.\ \ref{fig:snaps_overview}--\ref{fig:scheme_prol_spheroids}.

\begin{acknowledgments}
We thank Jens Bickmann, Julian Jeggle, and Fenna Stegemerten for helpful discussions. R.W.\ is funded by the Deutsche Forschungsgemeinschaft (DFG, German Research Foundation) -- 283183152 (WI 4170/3). The simulations for this work were performed on the computer cluster PALMA II of the University of M\"unster. 
\end{acknowledgments}

\appendix

\section{\label{sec:interactionpotential}Particle interactions}
We focus on hard particles and therefore use a short-ranged and purely repulsive interaction potential. More specifically, we use a modified version $u_\mathrm{int}$ of the Gay-Berne (GB) potential $u_\mathrm{GB}$~\cite{GayBerne1981}. The GB potential describes the interaction of ellipsoids, which depends on the relative position $\myvec{r}_{ij}= \myvec{r}_j - \myvec{r}_i$ of the interacting particles $i$ and $j$ as well as on their respective orientations $\myev{u}_i$ and $\myev{u}_j$.
It is repulsive for short distances and attractive for larger ones. Thus, we disregard the attractive part, such that the potential is purely repulsive. This can be done by \ZT{cutting} off the interaction potential at the minimum. In addition, we shift the potential to ensure that it is continuous and continuously differentiable (as done in Ref.\ \cite{JayaramFS2020}).
The GB potential reads~\cite{GayBerne1981}
\begin{equation}
\begin{split}
    &u_{\mathrm{GB}}({\myvec{r}_{ij}}, \myev{u}_i, \myev{u}_j) \\ &=4 \epsilon {(\epsilon_{\mathrm{GB}}^\prime (\myev{u}_i, \myev{u}_j))}^{\nu}
    {(\epsilon_{\mathrm{GB}}^\star (\myev{r}_{ij},\myev{u}_i, \myev{u}_j) )}^{\mu}
    \\&\bigg[ \bigg( \frac{1}{r_{ij}/\sigma_0 - \sigma_{\mathrm{GB},ij} +1} \bigg)^{12} - \bigg( \frac{1}{r_{ij}/\sigma_0 - \sigma_{\mathrm{GB},ij} +1} \bigg)^{6} \bigg] \raisetag{40pt}
\end{split}
\end{equation}
with the energy $\epsilon$ and the dimensionless energy parameters ${\epsilon_{\mathrm{GB}}^\prime }$ and ${\epsilon_{\mathrm{GB}}^\star  }$ of the potential, the exponents $\nu$ and $\mu$, the distance between the centers of the two particles $r_{ij} = \norm{\myvec{r}_{ij}}$, the length scale $\sigma_0$, and the distance of closest approach of the two particles $\sigma_{\mathrm{GB}, ij}$. 
Here, the energy parameter $\epsilon$ sets the energy of the potential.
The strength parameter $\epsilon_{\mathrm{GB}}^\prime $ results from the derivation of the overlap of two ellipsoids known as the \ZT{Gaussian overlap potential}~\cite{BerneP1972} and reads
\begin{align}
 \epsilon_{\mathrm{GB}}^\prime (\myev{u}_i, \myev{u}_j) = \big( 1 - \chi^2 (\myev{u}_i \cdot \myev{u}_j)^2 \big)^{-\frac{1}{2}} ,
\end{align}
where $\chi$ is the anisotropy parameter
\begin{align}
    \chi = \frac{\kappa^2-1}{\kappa^2+1}.
\end{align}
In the case of our ellipsoids, it is given by $\kappa = 2$. (For spheres, $\kappa$ equals one.) The exponent $\nu$ modifies the potential, while the other energy parameter $\epsilon_{\mathrm{GB}}^\star  $ allows to modify the interaction of the ellipsoids when two ellipsoids get close side-to-side versus end-to-end. Typically, the interaction is stronger when particles interact side-to-side, i.e., with their long sides close to each other, compared to particles interacting end-to-end, i.e., with their shortest sides close to each other. \\
This can be adjusted by the orientation-dependent energy parameter ${\epsilon_{\mathrm{GB}}^\star  }$, which was proposed by Gay and Berne~\cite{GayBerne1981} and reads
\begin{align}
    &{\epsilon_{\mathrm{GB}}^\star  } ( \myev{r}_{ij},\myev{u}_i, \myev{u}_j) \notag\\&= 1 - \frac{\chi^\prime}{2} \bigg[ \frac{ (\myev{r}_{ij} \cdot \myev{u}_i + \myev{r}_{ij} \cdot \myev{u}_j)^2  }   { 1 + \chi^\prime (\myev{u}_i \cdot \myev{u}_j)}  
    +
    \frac{ (\myev{r}_{ij} \cdot \myev{u}_i - \myev{r}_{ij} \cdot \myev{u}_j)^2  }   { 1 - \chi^\prime (\myev{u}_i \cdot \myev{u}_j)} 
      \bigg] \,  
\end{align}
with $\myev{r}_{ij} = \myvec{r}_{ij}/\norm{\myvec{r}_{ij}}$.
From the desired interaction strength for the side-to-side interaction $\epsilon_{\mathrm{side}}$ and the desired interaction strength for the end-to-end interaction $\epsilon_{\mathrm{end}}$, we obtain for the new parameter $\chi'$ the expression 
\begin{align}
    \chi^\prime = \frac{\epsilon_{\mathrm{side}}^{\frac{1}{\mu}} - \epsilon_{\mathrm{end}}^{\frac{1}{\mu}}}
            {\epsilon_{\mathrm{side}}^{\frac{1}{\mu}} + \epsilon_{\mathrm{end}}^{\frac{1}{\mu}}} \, 
\end{align}
that depends only on the relative strength $\epsilon_{\mathrm{side}}/\epsilon_{\mathrm{end}}$. 
The exponents $\nu$ and $\mu$ as well as the energy ratio $\epsilon_{\mathrm{side}}/\epsilon_{\mathrm{end}}$ can be chosen depending on the situation. Simulations of different particle shapes require different parameter combinations ~\cite{GayBerne1981, Rull1995, BerardiFZ1998}. 
In this work, we set $\epsilon_{\mathrm{side}}/\epsilon_{\mathrm{end}} = 1$, $\nu = 1$, and $\mu = 0$, so that our interaction potential is a Gaussian overlap potential. 

The distance of closest approach $\sigma_{\mathrm{GB},ij}$ can be calculated via
\begin{equation}
\begin{split}
    \sigma_{\mathrm{GB},ij}(\myev{r}_{ij},\myev{u}_i, \myev{u}_j)  &= 
    \bigg(  1 - \frac{\chi}{2}  \bigg[
    \frac{(\myev{r}_{ij} \cdot \myev{u}_i + \myev{r}_{ij} \cdot \myev{u}_2)^2  }
    { 1 + \chi (\myev{u}_i \cdot \myev{u}_j)  }\\
    &\qquad\qquad+
    \frac{(\myev{r}_{ij} \cdot \myev{u}_i - \myev{r}_{ij} \cdot \myev{u}_2)^2  }
    { 1 - \chi (\myev{u}_i \cdot \myev{u}_j)  }
    \bigg]  \bigg)^{-\frac{1}{2}} \,  .
\end{split}
\end{equation}
Note that $\sigma_{\mathrm{GB},ij}$ is not the exact distance of two ellipsoids, but a very common approximation. The exact way to calculate the distance between ellipsoids in two dimensions has only been discovered rather recently~\cite{ZhengP2007}, but it is computationally too expensive to be applicable in a large-scale computer simulation.  
As stated at the beginning of this section, our potential is a purely repulsive and short-ranged modification of the GB potential. 
This is achieved by cutting off the interaction at the minimum of the potential and shifting the potential. Following Refs.\ \cite{JayaramFS2020, BottWMSBW2018}, this yields 
\begin{equation}
 u_\mathrm{int}(\myvec{r}_{ij}, \myev{u}_i, \myev{u}_j ) = 
\begin{cases}
 u_\mathrm{GB}(\myvec{r}_{ij}, \myev{u}_i, \myev{u}_j ) + \epsilon_\mathrm{min}, & \mbox{if } r_{ij} \leq r_\mathrm{c}, \\
0, & \mbox{else}
\end{cases}
\label{eq:wca}%
\end{equation}
with $ \epsilon_\mathrm{min} =\epsilon {\epsilon_{\mathrm{GB}}^\prime }^{\nu}{\epsilon_{\mathrm{GB}}^\star  }^{\mu} $ and $r_\mathrm{c}= \sigma_0(2^{1/6} -1) + \sigma_{\mathrm{GB}, ij}$. 
The translational force resulting from the interaction is given by 
\begin{align}
 \myvec{F}_{\mathrm{int},i} (\lbrace \myvec{r}_j,  \myev{u}_j \rbrace   ) 
 = - \nabla_{\myvec{r}_i}   \sum_{j \neq i} u_{\mathrm{int}}( \myvec{r}_{ij}, \myev{u}_i, \myev{u}_j)
\end{align}
and the torque is
\begin{align}
M_{\mathrm{int},i} (\lbrace \myvec{r}_j,  \myev{u}_j \rbrace   ) = 
- \frac{\partial}{\partial \varphi_i}   \sum_{j \neq i} u_{\mathrm{int}}( \myvec{r}_{ij}, \myev{u}_i, \myev{u}_j) \, . 
\end{align}
\section{\label{sec:orderparameters}Calculation of polarization and nematic order}
We use the local polarization in the system as an order parameter. The polarization at an arbitrary position $\myvec{r}$ is calculated with the orientations of all nearby particles.
Each particle's orientation vector contributes to the polarization vector of that reference point, multiplied by a factor that depends on the distance between the particle and the reference point.
Therefore, the resulting polarization $\myvec{P}(\myvec{r})$ at a position $\myvec{r}$ reads
\begin{align}
    \myvec{P}(\myvec{r}) = \sum_{i=0}^{N} \myev{u}_i f_\mathrm{d}(\myvec{r},\myvec{r}_i,\myev{u}_i),
\label{eq:local_polarization}
\end{align}
where $N$ is the number of particles in the system, $\myvec{r}_i$ is the position of the $i$-th particle, $\myev{u}_i$ is the orientation of the $i$-th particle, and $f_\mathrm{d}$ is the distance-dependent scaling factor.
This factor ensures that only particles in the vicinity of $\myvec{r}$ contribute to the polarization measured at this point. This is because the particles' influence on the polarization decreases with a higher distance between the particle and the point of interest.

Since the scaling factor $f_\mathrm{d}$ depends on the distance of a particle to the reference point, we need a way to measure this distance.
For spheres, we can use the distance from the reference point to the particle's center.
However, in the case of ellipsoids the simple distance to the center of the ellipsoid should not be used as the particles are not spherical. Instead, considering the elliptic shape of the particles, prolate ellipsoidal coordinates are an excellent way to factor in their form.
The prolate spheroidal coordinate system is set up as follows: The center of the particle is the center of a new prolate spheroidal coordinate system, and the particle's orientation defines its $z$-axis. Three values define the new prolate spheroidal coordinate system: the distance parameter $\tau_\mathrm{prol} \geq 1$, the polar angular parameter $\zeta_\mathrm{prol}\in [-1,1]$, and the azimuthal angular parameter $\varphi_\mathrm{prol} \in [0, 2\pi [$. The azimuthal angular parameter can be dismissed as the ellipsoids only move in the two-dimensional plane here. 

\begin{figure}
    \centering
    \includegraphics{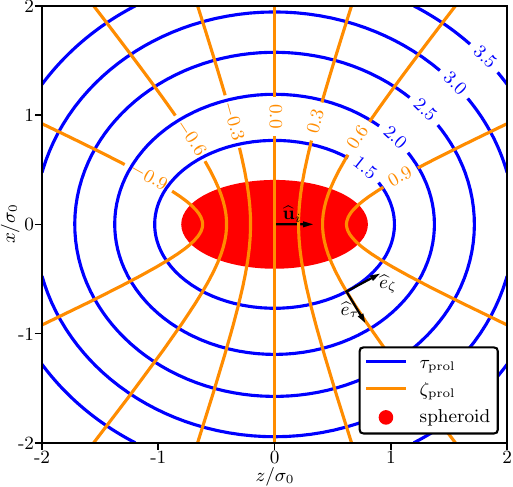}
    \caption{Scheme showing the curves representing prolate spheroidal coordinates for constant values of $\tau_\mathrm{prol}$ (blue) and $\zeta_\mathrm{prol}$ (orange) in the $z-x$-plane. The azimuthal angular parameter $\varphi_\mathrm{prol} $ induces rotation around the $z$-axis.}
    \label{fig:scheme_prol_spheroids}
\end{figure}

The values of the parameters can be calculated via 
\begin{align}
   \tau_\mathrm{prol} &= \frac{1}{2 d_\mathrm{foc}}
   \Big(\sqrt{{x^\prime}^2 + {y^\prime}^2 + (z^\prime+d_\mathrm{foc})^2 }\notag\\&  \quad + \sqrt{{x^\prime}^2 + {y^\prime}^2 + (z^\prime-d_\mathrm{foc})^2 } \Big) \, , \label{eq:tauproll}\\
    \zeta_\mathrm{prol} &= \frac{1}{2 d_\mathrm{foc}}
   \Big(\sqrt{{x^\prime}^2 + {y^\prime}^2 + (z^\prime+d_\mathrm{foc})^2 } \notag\\& \quad - \sqrt{{x^\prime}^2 + {y^\prime}^2 + (z^\prime-d_\mathrm{foc})^2 } \Big)  \, , \\
   \varphi_\mathrm{prol}  &= \arctan\! \bigg( \frac{y^\prime}{x^\prime} \bigg),
\end{align}
where $x^\prime$, $y^\prime$, and $z^\prime$ are the Cartesian coordinates of the reference point in the new reference frame and $2 d_\mathrm{foc} = \sqrt{(a_\mathrm{el}/2)^2 - (b_\mathrm{el}/2)^2}$ is the distance between the center of the ellipsoid and its focal point.
A scheme of the prolate coordinate system is shown in \cref{fig:scheme_prol_spheroids}. 
Varying the polar angular parameter $\zeta_\mathrm{prol}$ while keeping the distance parameter $\tau_\mathrm{prol}$ constant allows to draw prolate ellipsoids. The distance parameter $\tau_\mathrm{prol}$ is a good measurement for the distance from the ellipsoid.
We can now transfer the reference point into the new coordinate system, and the resulting value of $\tau_\mathrm{prol} $ is the distance.
Note that the value of $\tau_\mathrm{prol} $ is greater or equal to one.
Thus, we employ $\tau_\mathrm{prol} -1$ as the value for the distance from the center of the ellipsoid.
The scaling function is then defined as
\begin{equation}
\label{eq:distance_dependent_factor_pol}
f_\mathrm{d}(\tau_\mathrm{prol}) = \begin{cases}
\bar{f_\mathrm{d}}\exp\!\big(\frac{-1}{1-((\tau_\mathrm{prol}-1)/r_\mathrm{c})^2}\big), & \mbox{if } \tau_\mathrm{prol} < r_\mathrm{c}, \\
0, & \mbox{else,}
\end{cases}
\end{equation}
where $r_\mathrm{c} $ is the cut-off distance up to which a particle influences the polarization and $\bar{f_\mathrm{d}}$ is a normalization parameter. We found a good agreement between the visual inspection of the states and the polarization for $r_\mathrm{c} = 5\sigma_0 + a_\mathrm{el}/(2 d_\mathrm{foc}) $ for ellipsoids.
The normalization parameter $\bar{f_\mathrm{d}}$ fixes the volume.
Additionally, this parameter can be used to smear out the particles to determine a locally blurred density.
If the density were to be calculated, we would choose $\bar{f_\mathrm{d}}$ in such a way that we get the correct overall density, which means 
\begin{align}
\int_{A_\mathrm{part}}\!\!\!\!\!\! \mathrm{d}A \, H_\mathrm{part} = 
\int_{A_\mathrm{part}}\!\!\!\!\!\! \mathrm{d}A \, f_\mathrm{d}(\tau_\mathrm{prol}) 
\label{eq:heaviside}
\end{align}
where $A_\mathrm{part}$ is the area of the particle and $H_\mathrm{part}$ is a function that is one inside the particle and zero elsewhere. \Cref{eq:heaviside} defines $\bar{f_\mathrm{d}}$.
The function $f_\mathrm{d}$ is similar to a Gaussian distribution, but falls off to zero at a finite distance $\tau_\mathrm{prol} = r_\mathrm{c} $ while staying continuously differentiable.
Thus, the closer a particle is to the reference point, the stronger the orientation of the particle influences the polarization at the reference point.
By spreading a fine grid over the entire simulation domain and calculating the polarization for every grid point, we can calculate the average local polarization divided by density 
\begin{align}
\frac{\langle \norm{\myvec{P}}\rangle }{\Phi_0} = \frac{1}{N_\mathrm{grid} \Phi_0} \sum_{\myvec{r}_i \in {R}_\mathrm{grid}} 
  \norm{\myvec{P}(\myvec{r}_i)},
\end{align}
where ${R}_\mathrm{grid}$ denotes the set of all grid points and $N_\mathrm{grid}$ denotes the number of all grid points. We chose a distance of $0.1 \sigma_0$ between two grid points.
For each grid point, the length of the polarization vector is calculated. Then, we average over the resulting values.
Applying this method, we measure the average local polarization $\langle \norm{\myvec{P}}\rangle /\Phi_0$ for a single time and the whole simulation domain.
Additionally, we can average over 300 time steps to get a reliable measurement of the typical polarization for a given parameter set. 
Thereby, the values of the polarization obtained for different system parameters (\Pe{} number and packing density) can be compared.
Thus, the average local polarization we use to characterize a system is averaged over the simulation domain and over time. 
We can calculate the average global polarization
\begin{equation}
 \frac{\norm{\langle\myvec{P}\rangle } }{\Phi_0} = \frac{1}{N_\mathrm{grid} \Phi_0}\bignorm{ \sum_{\myvec{r}_i \in {R}_\mathrm{grid}} 
  \myvec{P}(\myvec{r}_i)}
\end{equation}
of a system by averaging over the local polarization vector grid (a square lattice), calculating the norm of the resulting vector, and dividing by $\Phi_0$. By averaging over the local polarization vectors instead of their norm, we can measure whether the whole system is polarized.  

In addition to the polarization, the average nematic order divided by density $\langle S \rangle /\Phi_0$ can be used as an order parameter. It is calculated via the nematic tensor $\mymat{Q}$, which is given by~\cite{ShankarSBMV2022topological} 
\begin{align}
    \mymat{Q}(\myvec{r}) &= \sum_{i=0}^{N} 
    (\myev{u}_i f_\mathrm{d}(\myvec{r},\myvec{r}_i,\myev{u}_i) \otimes 
     \myev{u}_i f_\mathrm{d}(\myvec{r},\myvec{r}_i,\myev{u}_i)) - \frac{\mymat{1} }{2}.
\end{align}
From the nematic tensor $\mymat{Q}$, we can calculate the nematic order $S$ using
\begin{align}
    S^2(\myvec{r}) = 2 \mathrm{Tr}(\mymat{Q}^2(\myvec{r}) )
\end{align}
with the trace $\mathrm{Tr}$. This is done for every point of a fine grid over the entire simulation domain. Then, we average over these points and time. As a last step, the result is divided by density. This gives the average nematic order divided by density, which is defined as
\begin{align}
\frac{\langle S \rangle }{\Phi_0} = \frac{1}{N_\mathrm{grid} \Phi_0} \sum_{\myvec{r}_i \in {R}_\mathrm{grid}} 
  \sqrt{S^2(\myvec{r}_i)} \, .
\end{align}

\section{\label{sec:tables}Tables of fit parameters}
These tables contain the fit parameters required for the function $h(\mathrm{Pe}, \Phi_0)$ defined in \cref{eq:fitfunc_PePhi}.
\begin{table*}[!htbp]
    \fontsize{9}{0.0}
    \setlength{\tabcolsep}{1pt}
    \renewcommand{\arraystretch}{0.3}
    \centering
\begin{tabularx}{\linewidth}{|c|c|Y|Y|Y|Y|Y|Y|Y|}
      \cline{3-9} 
       \multicolumn{2}{c|}{} & & & & & & & \\
      \multicolumn{2}{c|}{} & $\hparamvalue_{-2,0}$  & $\hparamvalue_{-2, 1}$ & $\hparamvalue_{-2, 2}$ & $\hparamvalue_{ -2, 3}$ & $\hparamvalue_{-1 ,0 }$ & $\hparamvalue_{-1 ,1 }$ &$\hparamvalue_{-1 ,2 }$  \\
      \multicolumn{2}{c|}{} & & & & & & & \\
      \multicolumn{2}{c|}{} & $\hparamvalue_{ -1, 3}$& $\hparamvalue_{ 0, 0}$ & $\hparamvalue_{ 0,1 }$ & $\hparamvalue_{0 ,2 }$ & $\hparamvalue_{0 , 3}$ & $\hparamvalue_{1 ,0 }$ &$\hparamvalue_{1 ,1 }$  \\
      \multicolumn{2}{c|}{} & & & & & & & \\
      \multicolumn{2}{c|}{} & $\hparamvalue_{ 1,2 }$  & $\hparamvalue_{1 ,3 }$ & $\hparamvalue_{ 2,0 }$ & $\hparamvalue_{ 2, 1}$ & $\hparamvalue_{ 2,2 }$ & $\hparamvalue_{2 ,3 }$ & \\    
      \multicolumn{2}{c|}{} & & & & & & & \\\hline
& & & & & & & & \\
\hline
\end{tabularx}
\caption{\label{tab:param_table_1}Fit coefficients of the function $h(\mathrm{Pe}, \Phi_0)$ used to fit the variables of the Fourier coefficients $a_{0,0}^{1, 1}$ and $a_{0,1}^{1, 1}$ of the function $g_\mathrm{approx}$.}
\end{table*}

\begin{table*}[!htbp]
    \fontsize{9}{0.0}
    \setlength{\tabcolsep}{1pt}
    \renewcommand{\arraystretch}{0.3}
    \centering
    \begin{tabularx}{\linewidth}{|c|c|Y|Y|Y|Y|Y|Y|Y|}
      \cline{3-9} 
       \multicolumn{2}{c|}{} & & & & & & & \\
      \multicolumn{2}{c|}{} & $\hparamvalue_{-2,0}$  & $\hparamvalue_{-2, 1}$ & $\hparamvalue_{-2, 2}$ & $\hparamvalue_{ -2, 3}$ & $\hparamvalue_{-1 ,0 }$ & $\hparamvalue_{-1 ,1 }$ &$\hparamvalue_{-1 ,2 }$  \\
      \multicolumn{2}{c|}{} & & & & & & & \\
      \multicolumn{2}{c|}{} & $\hparamvalue_{ -1, 3}$& $\hparamvalue_{ 0, 0}$ & $\hparamvalue_{ 0,1 }$ & $\hparamvalue_{0 ,2 }$ & $\hparamvalue_{0 , 3}$ & $\hparamvalue_{1 ,0 }$ &$\hparamvalue_{1 ,1 }$  \\
      \multicolumn{2}{c|}{} & & & & & & & \\
      \multicolumn{2}{c|}{} & $\hparamvalue_{ 1,2 }$  & $\hparamvalue_{1 ,3 }$ & $\hparamvalue_{ 2,0 }$ & $\hparamvalue_{ 2, 1}$ & $\hparamvalue_{ 2,2 }$ & $\hparamvalue_{2 ,3 }$ & \\    
      \multicolumn{2}{c|}{} & & & & & & & \\\hline
& & & & & & & & \\
\hline
\end{tabularx}
\caption{\label{tab:param_table_2}Fit coefficients of the function $h(\mathrm{Pe}, \Phi_0)$ used to fit the variables of the Fourier coefficients $a_{0,2}^{1, 1}$ and $a_{0,3}^{1, 1}$ of the function $g_\mathrm{approx}$.}

\end{table*}

\begin{table*}[!htbp]
    \fontsize{9}{0.0}
    \setlength{\tabcolsep}{1pt}
    \renewcommand{\arraystretch}{0.3}
    \centering
    \begin{tabularx}{\linewidth}{|c|c|Y|Y|Y|Y|Y|Y|Y|}
      \cline{3-9} 
       \multicolumn{2}{c|}{} & & & & & & & \\
      \multicolumn{2}{c|}{} & $\hparamvalue_{-2,0}$  & $\hparamvalue_{-2, 1}$ & $\hparamvalue_{-2, 2}$ & $\hparamvalue_{ -2, 3}$ & $\hparamvalue_{-1 ,0 }$ & $\hparamvalue_{-1 ,1 }$ &$\hparamvalue_{-1 ,2 }$  \\
      \multicolumn{2}{c|}{} & & & & & & & \\
      \multicolumn{2}{c|}{} & $\hparamvalue_{ -1, 3}$& $\hparamvalue_{ 0, 0}$ & $\hparamvalue_{ 0,1 }$ & $\hparamvalue_{0 ,2 }$ & $\hparamvalue_{0 , 3}$ & $\hparamvalue_{1 ,0 }$ &$\hparamvalue_{1 ,1 }$  \\
      \multicolumn{2}{c|}{} & & & & & & & \\
      \multicolumn{2}{c|}{} & $\hparamvalue_{ 1,2 }$  & $\hparamvalue_{1 ,3 }$ & $\hparamvalue_{ 2,0 }$ & $\hparamvalue_{ 2, 1}$ & $\hparamvalue_{ 2,2 }$ & $\hparamvalue_{2 ,3 }$ & \\    
      \multicolumn{2}{c|}{} & & & & & & & \\\hline
& & & & & & & & \\
\hline
\end{tabularx}
\caption{\label{tab:param_table_3}Fit coefficients of the function $h(\mathrm{Pe}, \Phi_0)$ used to fit the variables of the Fourier coefficients $a_{1,0}^{1, 1}$ and $a_{1,1}^{1, 1}$ of the function $g_\mathrm{approx}$.}

\end{table*}

\begin{table*}[!htbp]
    \fontsize{9}{0.0}
    \setlength{\tabcolsep}{1pt}
    \renewcommand{\arraystretch}{0.3}
    \centering
    \begin{tabularx}{\linewidth}{|c|c|Y|Y|Y|Y|Y|Y|Y|}
      \cline{3-9} 
       \multicolumn{2}{c|}{} & & & & & & & \\
      \multicolumn{2}{c|}{} & $\hparamvalue_{-2,0}$  & $\hparamvalue_{-2, 1}$ & $\hparamvalue_{-2, 2}$ & $\hparamvalue_{ -2, 3}$ & $\hparamvalue_{-1 ,0 }$ & $\hparamvalue_{-1 ,1 }$ &$\hparamvalue_{-1 ,2 }$  \\
      \multicolumn{2}{c|}{} & & & & & & & \\
      \multicolumn{2}{c|}{} & $\hparamvalue_{ -1, 3}$& $\hparamvalue_{ 0, 0}$ & $\hparamvalue_{ 0,1 }$ & $\hparamvalue_{0 ,2 }$ & $\hparamvalue_{0 , 3}$ & $\hparamvalue_{1 ,0 }$ &$\hparamvalue_{1 ,1 }$  \\
      \multicolumn{2}{c|}{} & & & & & & & \\
      \multicolumn{2}{c|}{} & $\hparamvalue_{ 1,2 }$  & $\hparamvalue_{1 ,3 }$ & $\hparamvalue_{ 2,0 }$ & $\hparamvalue_{ 2, 1}$ & $\hparamvalue_{ 2,2 }$ & $\hparamvalue_{2 ,3 }$ & \\    
      \multicolumn{2}{c|}{} & & & & & & & \\\hline
& & & & & & & & \\
\hline
\end{tabularx}
\caption{\label{tab:param_table_4}Fit coefficients of the function $h(\mathrm{Pe}, \Phi_0)$ used to fit the variables of the Fourier coefficients $a_{1,2}^{1, 1}$ and $a_{1,3}^{1, 1}$ of the function $g_\mathrm{approx}$.}

\end{table*}

\begin{table*}[!htbp]
    \fontsize{9}{0.0}
    \setlength{\tabcolsep}{1pt}
    \renewcommand{\arraystretch}{0.3}
    \centering
    \begin{tabularx}{\linewidth}{|c|c|Y|Y|Y|Y|Y|Y|Y|}
      \cline{3-9} 
       \multicolumn{2}{c|}{} & & & & & & & \\
      \multicolumn{2}{c|}{} & $\hparamvalue_{-2,0}$  & $\hparamvalue_{-2, 1}$ & $\hparamvalue_{-2, 2}$ & $\hparamvalue_{ -2, 3}$ & $\hparamvalue_{-1 ,0 }$ & $\hparamvalue_{-1 ,1 }$ &$\hparamvalue_{-1 ,2 }$  \\
      \multicolumn{2}{c|}{} & & & & & & & \\
      \multicolumn{2}{c|}{} & $\hparamvalue_{ -1, 3}$& $\hparamvalue_{ 0, 0}$ & $\hparamvalue_{ 0,1 }$ & $\hparamvalue_{0 ,2 }$ & $\hparamvalue_{0 , 3}$ & $\hparamvalue_{1 ,0 }$ &$\hparamvalue_{1 ,1 }$  \\
      \multicolumn{2}{c|}{} & & & & & & & \\
      \multicolumn{2}{c|}{} & $\hparamvalue_{ 1,2 }$  & $\hparamvalue_{1 ,3 }$ & $\hparamvalue_{ 2,0 }$ & $\hparamvalue_{ 2, 1}$ & $\hparamvalue_{ 2,2 }$ & $\hparamvalue_{2 ,3 }$ & \\    
      \multicolumn{2}{c|}{} & & & & & & & \\\hline
& & & & & & & & \\
\hline
\end{tabularx}
\caption{\label{tab:param_table_5}Fit coefficients of the function $h(\mathrm{Pe}, \Phi_0)$ used to fit the variables of the Fourier coefficients $a_{2,0}^{1, 1}$ and $a_{2,1}^{1, 1}$ of the function $g_\mathrm{approx}$.}

\end{table*}

\begin{table*}[!htbp]
    \fontsize{9}{0.0}
    \setlength{\tabcolsep}{1pt}
    \renewcommand{\arraystretch}{0.3}
    \centering
    \begin{tabularx}{\linewidth}{|c|c|Y|Y|Y|Y|Y|Y|Y|}
      \cline{3-9} 
       \multicolumn{2}{c|}{} & & & & & & & \\
      \multicolumn{2}{c|}{} & $\hparamvalue_{-2,0}$  & $\hparamvalue_{-2, 1}$ & $\hparamvalue_{-2, 2}$ & $\hparamvalue_{ -2, 3}$ & $\hparamvalue_{-1 ,0 }$ & $\hparamvalue_{-1 ,1 }$ &$\hparamvalue_{-1 ,2 }$  \\
      \multicolumn{2}{c|}{} & & & & & & & \\
      \multicolumn{2}{c|}{} & $\hparamvalue_{ -1, 3}$& $\hparamvalue_{ 0, 0}$ & $\hparamvalue_{ 0,1 }$ & $\hparamvalue_{0 ,2 }$ & $\hparamvalue_{0 , 3}$ & $\hparamvalue_{1 ,0 }$ &$\hparamvalue_{1 ,1 }$  \\
      \multicolumn{2}{c|}{} & & & & & & & \\
      \multicolumn{2}{c|}{} & $\hparamvalue_{ 1,2 }$  & $\hparamvalue_{1 ,3 }$ & $\hparamvalue_{ 2,0 }$ & $\hparamvalue_{ 2, 1}$ & $\hparamvalue_{ 2,2 }$ & $\hparamvalue_{2 ,3 }$ & \\    
      \multicolumn{2}{c|}{} & & & & & & & \\\hline
& & & & & & & & \\
\hline
\end{tabularx}
\caption{\label{tab:param_table_6}Fit coefficients of the function $h(\mathrm{Pe}, \Phi_0)$ used to fit the variables of the Fourier coefficients $a_{2,2}^{1, 1}$ and $a_{2,3}^{1, 1}$ of the function $g_\mathrm{approx}$.}

\end{table*}

\begin{table*}[!htbp]
    \fontsize{9}{0.0}
    \setlength{\tabcolsep}{1pt}
    \renewcommand{\arraystretch}{0.3}
    \centering
    \begin{tabularx}{\linewidth}{|c|c|Y|Y|Y|Y|Y|Y|Y|}
      \cline{3-9} 
       \multicolumn{2}{c|}{} & & & & & & & \\
      \multicolumn{2}{c|}{} & $\hparamvalue_{-2,0}$  & $\hparamvalue_{-2, 1}$ & $\hparamvalue_{-2, 2}$ & $\hparamvalue_{ -2, 3}$ & $\hparamvalue_{-1 ,0 }$ & $\hparamvalue_{-1 ,1 }$ &$\hparamvalue_{-1 ,2 }$  \\
      \multicolumn{2}{c|}{} & & & & & & & \\
      \multicolumn{2}{c|}{} & $\hparamvalue_{ -1, 3}$& $\hparamvalue_{ 0, 0}$ & $\hparamvalue_{ 0,1 }$ & $\hparamvalue_{0 ,2 }$ & $\hparamvalue_{0 , 3}$ & $\hparamvalue_{1 ,0 }$ &$\hparamvalue_{1 ,1 }$  \\
      \multicolumn{2}{c|}{} & & & & & & & \\
      \multicolumn{2}{c|}{} & $\hparamvalue_{ 1,2 }$  & $\hparamvalue_{1 ,3 }$ & $\hparamvalue_{ 2,0 }$ & $\hparamvalue_{ 2, 1}$ & $\hparamvalue_{ 2,2 }$ & $\hparamvalue_{2 ,3 }$ & \\    
      \multicolumn{2}{c|}{} & & & & & & & \\\hline
& & & & & & & & \\
\hline
\end{tabularx}
\caption{\label{tab:param_table_7}Fit coefficients of the function $h(\mathrm{Pe}, \Phi_0)$ used to fit the variables of the Fourier coefficients $a_{3,0}^{1, 1}$ and $a_{3,1}^{1, 1}$ of the function $g_\mathrm{approx}$.}

\end{table*}

\begin{table*}[!htbp]
    \fontsize{9}{0.0}
    \setlength{\tabcolsep}{1pt}
    \renewcommand{\arraystretch}{0.3}
    \centering
    \begin{tabularx}{\linewidth}{|c|c|Y|Y|Y|Y|Y|Y|Y|}
      \cline{3-9} 
       \multicolumn{2}{c|}{} & & & & & & & \\
      \multicolumn{2}{c|}{} & $\hparamvalue_{-2,0}$  & $\hparamvalue_{-2, 1}$ & $\hparamvalue_{-2, 2}$ & $\hparamvalue_{ -2, 3}$ & $\hparamvalue_{-1 ,0 }$ & $\hparamvalue_{-1 ,1 }$ &$\hparamvalue_{-1 ,2 }$  \\
      \multicolumn{2}{c|}{} & & & & & & & \\
      \multicolumn{2}{c|}{} & $\hparamvalue_{ -1, 3}$& $\hparamvalue_{ 0, 0}$ & $\hparamvalue_{ 0,1 }$ & $\hparamvalue_{0 ,2 }$ & $\hparamvalue_{0 , 3}$ & $\hparamvalue_{1 ,0 }$ &$\hparamvalue_{1 ,1 }$  \\
      \multicolumn{2}{c|}{} & & & & & & & \\
      \multicolumn{2}{c|}{} & $\hparamvalue_{ 1,2 }$  & $\hparamvalue_{1 ,3 }$ & $\hparamvalue_{ 2,0 }$ & $\hparamvalue_{ 2, 1}$ & $\hparamvalue_{ 2,2 }$ & $\hparamvalue_{2 ,3 }$ & \\    
      \multicolumn{2}{c|}{} & & & & & & & \\\hline
& & & & & & & & \\
\hline
\end{tabularx}
\caption{\label{tab:param_table_8}Fit coefficients of the function $h(\mathrm{Pe}, \Phi_0)$ used to fit the variables of the Fourier coefficients $a_{3,2}^{1, 1}$ and $a_{3,3}^{1, 1}$ of the function $g_\mathrm{approx}$.}

\end{table*}

\begin{table*}[!htbp]
    \fontsize{9}{0.0}
    \setlength{\tabcolsep}{1pt}
    \renewcommand{\arraystretch}{0.3}
    \centering
    \begin{tabularx}{\linewidth}{|c|c|Y|Y|Y|Y|Y|Y|Y|}
      \cline{3-9} 
       \multicolumn{2}{c|}{} & & & & & & & \\
      \multicolumn{2}{c|}{} & $\hparamvalue_{-2,0}$  & $\hparamvalue_{-2, 1}$ & $\hparamvalue_{-2, 2}$ & $\hparamvalue_{ -2, 3}$ & $\hparamvalue_{-1 ,0 }$ & $\hparamvalue_{-1 ,1 }$ &$\hparamvalue_{-1 ,2 }$  \\
      \multicolumn{2}{c|}{} & & & & & & & \\
      \multicolumn{2}{c|}{} & $\hparamvalue_{ -1, 3}$& $\hparamvalue_{ 0, 0}$ & $\hparamvalue_{ 0,1 }$ & $\hparamvalue_{0 ,2 }$ & $\hparamvalue_{0 , 3}$ & $\hparamvalue_{1 ,0 }$ &$\hparamvalue_{1 ,1 }$  \\
      \multicolumn{2}{c|}{} & & & & & & & \\
      \multicolumn{2}{c|}{} & $\hparamvalue_{ 1,2 }$  & $\hparamvalue_{1 ,3 }$ & $\hparamvalue_{ 2,0 }$ & $\hparamvalue_{ 2, 1}$ & $\hparamvalue_{ 2,2 }$ & $\hparamvalue_{2 ,3 }$ & \\    
      \multicolumn{2}{c|}{} & & & & & & & \\\hline
& & & & & & & & \\
\hline
\end{tabularx}
\caption{\label{tab:param_table_9}Fit coefficients of the function $h(\mathrm{Pe}, \Phi_0)$ used to fit the variables of the Fourier coefficients $a_{1,1}^{2, 2}$ and $a_{1,2}^{2, 2}$ of the function $g_\mathrm{approx}$.}

\end{table*}

\begin{table*}[!htbp]
    \fontsize{9}{0.0}
    \setlength{\tabcolsep}{1pt}
    \renewcommand{\arraystretch}{0.3}
    \centering
    \begin{tabularx}{\linewidth}{|c|c|Y|Y|Y|Y|Y|Y|Y|}
      \cline{3-9} 
       \multicolumn{2}{c|}{} & & & & & & & \\
      \multicolumn{2}{c|}{} & $\hparamvalue_{-2,0}$  & $\hparamvalue_{-2, 1}$ & $\hparamvalue_{-2, 2}$ & $\hparamvalue_{ -2, 3}$ & $\hparamvalue_{-1 ,0 }$ & $\hparamvalue_{-1 ,1 }$ &$\hparamvalue_{-1 ,2 }$  \\
      \multicolumn{2}{c|}{} & & & & & & & \\
      \multicolumn{2}{c|}{} & $\hparamvalue_{ -1, 3}$& $\hparamvalue_{ 0, 0}$ & $\hparamvalue_{ 0,1 }$ & $\hparamvalue_{0 ,2 }$ & $\hparamvalue_{0 , 3}$ & $\hparamvalue_{1 ,0 }$ &$\hparamvalue_{1 ,1 }$  \\
      \multicolumn{2}{c|}{} & & & & & & & \\
      \multicolumn{2}{c|}{} & $\hparamvalue_{ 1,2 }$  & $\hparamvalue_{1 ,3 }$ & $\hparamvalue_{ 2,0 }$ & $\hparamvalue_{ 2, 1}$ & $\hparamvalue_{ 2,2 }$ & $\hparamvalue_{2 ,3 }$ & \\    
      \multicolumn{2}{c|}{} & & & & & & & \\\hline
& & & & & & & & \\
\hline
\end{tabularx}
\caption{\label{tab:param_table_10}Fit coefficients of the function $h(\mathrm{Pe}, \Phi_0)$ used to fit the variables of the Fourier coefficients $a_{1,3}^{2, 2}$ and $a_{2,1}^{2, 2}$ of the function $g_\mathrm{approx}$.}

\end{table*}

\begin{table*}[!htbp]
    \fontsize{9}{0.0}
    \setlength{\tabcolsep}{1pt}
    \renewcommand{\arraystretch}{0.3}
    \centering
    \begin{tabularx}{\linewidth}{|c|c|Y|Y|Y|Y|Y|Y|Y|}
      \cline{3-9} 
       \multicolumn{2}{c|}{} & & & & & & & \\
      \multicolumn{2}{c|}{} & $\hparamvalue_{-2,0}$  & $\hparamvalue_{-2, 1}$ & $\hparamvalue_{-2, 2}$ & $\hparamvalue_{ -2, 3}$ & $\hparamvalue_{-1 ,0 }$ & $\hparamvalue_{-1 ,1 }$ &$\hparamvalue_{-1 ,2 }$  \\
      \multicolumn{2}{c|}{} & & & & & & & \\
      \multicolumn{2}{c|}{} & $\hparamvalue_{ -1, 3}$& $\hparamvalue_{ 0, 0}$ & $\hparamvalue_{ 0,1 }$ & $\hparamvalue_{0 ,2 }$ & $\hparamvalue_{0 , 3}$ & $\hparamvalue_{1 ,0 }$ &$\hparamvalue_{1 ,1 }$  \\
      \multicolumn{2}{c|}{} & & & & & & & \\
      \multicolumn{2}{c|}{} & $\hparamvalue_{ 1,2 }$  & $\hparamvalue_{1 ,3 }$ & $\hparamvalue_{ 2,0 }$ & $\hparamvalue_{ 2, 1}$ & $\hparamvalue_{ 2,2 }$ & $\hparamvalue_{2 ,3 }$ & \\    
      \multicolumn{2}{c|}{} & & & & & & & \\\hline
& & & & & & & & \\
\hline
\end{tabularx}
\caption{\label{tab:param_table_11}Fit coefficients of the function $h(\mathrm{Pe}, \Phi_0)$ used to fit the variables of the Fourier coefficients $a_{2,2}^{2, 2}$ and $a_{2,3}^{2, 2}$ of the function $g_\mathrm{approx}$.}

\end{table*}

\begin{table*}[!htbp]
    \fontsize{9}{0.0}
    \setlength{\tabcolsep}{1pt}
    \renewcommand{\arraystretch}{0.3}
    \centering
    \begin{tabularx}{\linewidth}{|c|c|Y|Y|Y|Y|Y|Y|Y|}
      \cline{3-9} 
       \multicolumn{2}{c|}{} & & & & & & & \\
      \multicolumn{2}{c|}{} & $\hparamvalue_{-2,0}$  & $\hparamvalue_{-2, 1}$ & $\hparamvalue_{-2, 2}$ & $\hparamvalue_{ -2, 3}$ & $\hparamvalue_{-1 ,0 }$ & $\hparamvalue_{-1 ,1 }$ &$\hparamvalue_{-1 ,2 }$  \\
      \multicolumn{2}{c|}{} & & & & & & & \\
      \multicolumn{2}{c|}{} & $\hparamvalue_{ -1, 3}$& $\hparamvalue_{ 0, 0}$ & $\hparamvalue_{ 0,1 }$ & $\hparamvalue_{0 ,2 }$ & $\hparamvalue_{0 , 3}$ & $\hparamvalue_{1 ,0 }$ &$\hparamvalue_{1 ,1 }$  \\
      \multicolumn{2}{c|}{} & & & & & & & \\
      \multicolumn{2}{c|}{} & $\hparamvalue_{ 1,2 }$  & $\hparamvalue_{1 ,3 }$ & $\hparamvalue_{ 2,0 }$ & $\hparamvalue_{ 2, 1}$ & $\hparamvalue_{ 2,2 }$ & $\hparamvalue_{2 ,3 }$ & \\    
      \multicolumn{2}{c|}{} & & & & & & & \\\hline
& & & & & & & & \\
\hline
\end{tabularx}
\caption{\label{tab:param_table_12}Fit coefficients of the function $h(\mathrm{Pe}, \Phi_0)$ used to fit the variables of the Fourier coefficients $a_{3,1}^{2, 2}$ and $a_{3,2}^{2, 2}$ of the function $g_\mathrm{approx}$.}

\end{table*}

\begin{table*}[!htbp]
    \fontsize{9}{0.0}
    \setlength{\tabcolsep}{1pt}
    \renewcommand{\arraystretch}{0.3}
    \centering
    \begin{tabularx}{\linewidth}{|c|c|Y|Y|Y|Y|Y|Y|Y|}
      \cline{3-9} 
       \multicolumn{2}{c|}{} & & & & & & & \\
      \multicolumn{2}{c|}{} & $\hparamvalue_{-2,0}$  & $\hparamvalue_{-2, 1}$ & $\hparamvalue_{-2, 2}$ & $\hparamvalue_{ -2, 3}$ & $\hparamvalue_{-1 ,0 }$ & $\hparamvalue_{-1 ,1 }$ &$\hparamvalue_{-1 ,2 }$  \\
      \multicolumn{2}{c|}{} & & & & & & & \\
      \multicolumn{2}{c|}{} & $\hparamvalue_{ -1, 3}$& $\hparamvalue_{ 0, 0}$ & $\hparamvalue_{ 0,1 }$ & $\hparamvalue_{0 ,2 }$ & $\hparamvalue_{0 , 3}$ & $\hparamvalue_{1 ,0 }$ &$\hparamvalue_{1 ,1 }$  \\
      \multicolumn{2}{c|}{} & & & & & & & \\
      \multicolumn{2}{c|}{} & $\hparamvalue_{ 1,2 }$  & $\hparamvalue_{1 ,3 }$ & $\hparamvalue_{ 2,0 }$ & $\hparamvalue_{ 2, 1}$ & $\hparamvalue_{ 2,2 }$ & $\hparamvalue_{2 ,3 }$ & \\    
      \multicolumn{2}{c|}{} & & & & & & & \\\hline
& & & & & & & & \\
\hline
\end{tabularx}
\caption{\label{tab:param_table_13}Fit coefficients of the function $h(\mathrm{Pe}, \Phi_0)$ used to fit the variables of the Fourier coefficient $a_{3,3}^{2, 2}$ of the function $g_\mathrm{approx}$.}

\end{table*}

\bibliography{refs,control}
\end{document}